\documentclass[preprint,12pt]{elsarticle}

\usepackage{amssymb}
\usepackage{amsmath}
\usepackage{xcolor}
\usepackage{hyperref}       
\usepackage{url}            
\usepackage{booktabs}       
\usepackage{amsfonts}       
\usepackage{nicefrac}       
\usepackage{microtype}      
\usepackage{lipsum}
\usepackage{float}
\usepackage{fancyhdr}       
\usepackage{graphicx}       
\usepackage{subcaption}
\usepackage{listings}
\usepackage[most]{tcolorbox}
\usepackage[normalem]{ulem} 

\usepackage{mathrsfs}

\microtypesetup{expansion=false}

\begin{document}

\begin{frontmatter}




\title{
Algorithmic Control Improves Residential Building Energy and EV Management when PV Capacity is High but Battery Capacity is Low
}


 \author[label1]{Lennart Ullner}
\affiliation[label1]{organization={Humboldt University of Berlin},
            country={Germany}}

 \author[label1]{Alona Zharova
 \corref{cor1}\texorpdfstring{\href{https://orcid.org/0000-0003-3506-4744}
 {\includegraphics[scale=0.06]{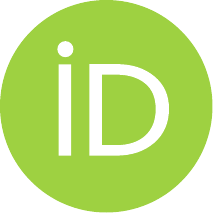}}}{}}
 \ead{alona.zharova@hu-berlin.de}
 \cortext[cor1]{Correspondence author}

 \author[label3,label4]{Felix Creutzig
 \corref{cor2}\texorpdfstring{\href{https://orcid.org/0000-0002-5710-3348}
 {\includegraphics[scale=0.06]{Visualizations/orcid.pdf}}}{}}
\affiliation[label3]{organization={Potsdam Institute for Climate Impact Research},
            country={Germany}}

\affiliation[label4]{organization={Bennett Institute for Innovation and Policy Acceleration, University of Sussex, Falmer},
            country={United Kingdom}}


\begin{abstract}
Efficient energy management in prosumer households is key to alleviating grid stress in an energy transition marked by electric vehicles (EV), renewable energies and battery storage.
However, it is unclear how households optimize prosumer EV charging. 
Here we study real-world data from 90 households on fixed-rate electricity tariffs in German-speaking countries to investigate the potential of Deep Reinforcement Learning (DRL) and other control approaches (Rule-Based, Model Predictive Control) to manage the dynamic and uncertain environment of Home Energy Management (HEM) and optimize household charging patterns. 
The DRL agent efficiently aligns charging of EV and battery storage with PV surplus. 
We find that frequent EV charging transactions, early EV connections and PV surplus increase optimization potential. 
A detailed analysis of nine households (1 hour resolution, 1 year) demonstrates that high battery capacity facilitates self optimization; in this case further algorithmic control shows little value. In cases with relatively low battery capacity, algorithmic control with DRL improves energy management and cost savings by a relevant margin. 
This result is further corroborated by our simulation of a synthetic household.
We conclude that prosumer households with optimization potential would profit from DRL, thus benefiting also the full electricity system and its decarbonization.


\end{abstract}


\begin{graphicalabstract}
    \centering
    \includegraphics[width=0.7\linewidth]{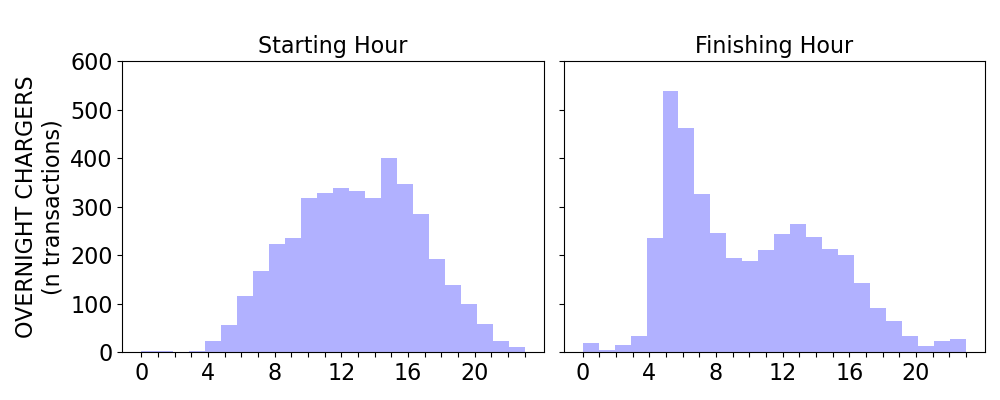}

    \vspace{-0.25cm}
    
    \includegraphics[width=0.7\linewidth]{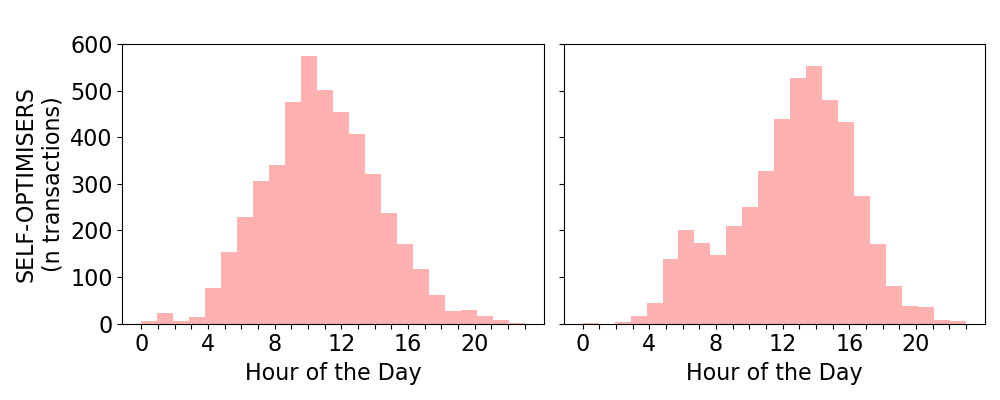} 


    \includegraphics[width=0.7\linewidth]{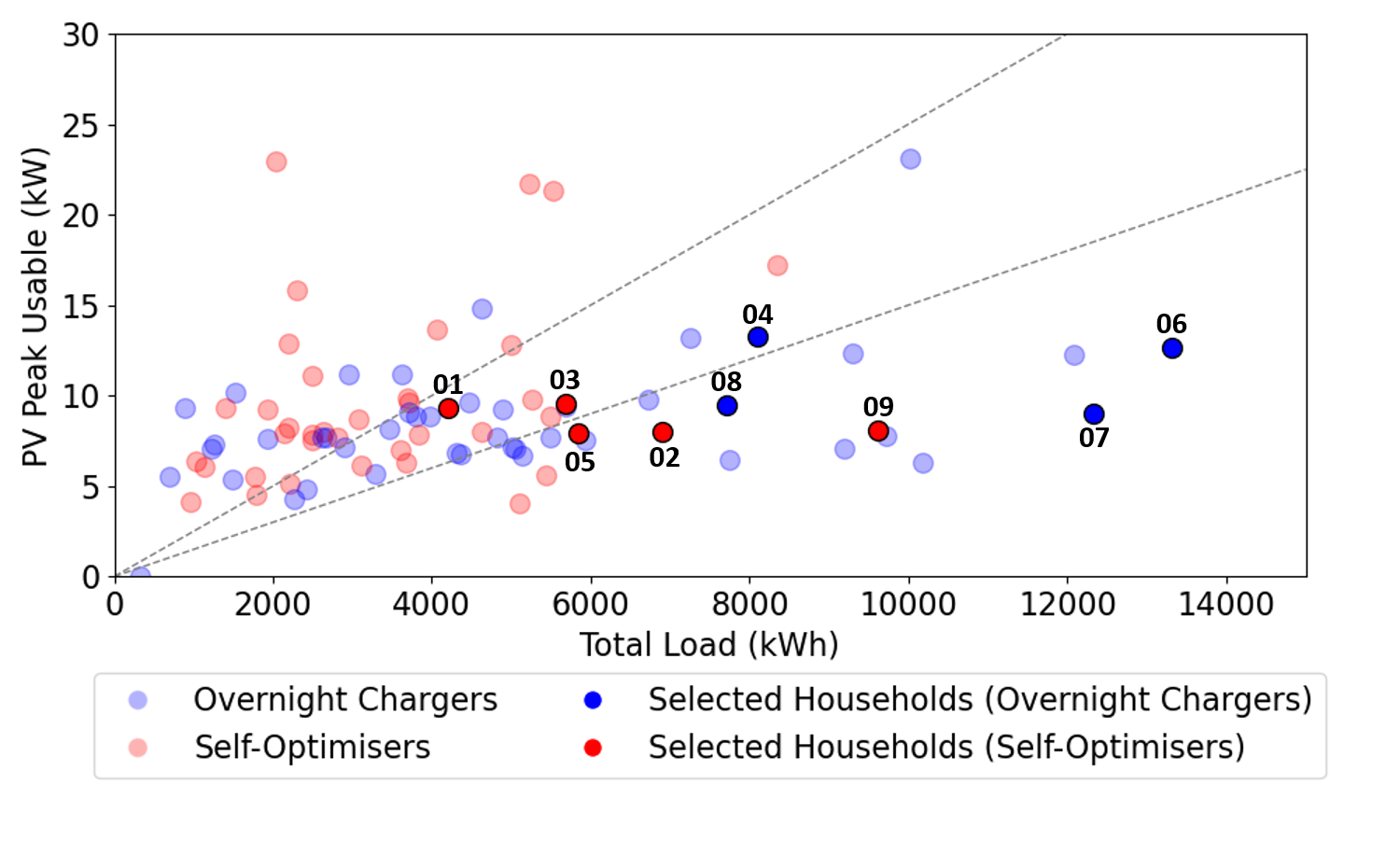}

    
    \label{fig:TransactionTimesHistogram}
    
\end{graphicalabstract}

\begin{highlights}
    \item DRL algorithm with DDPG effectively aligns charging of EV and BESS with PV surplus 
    \item DRL performance is context-dependent and requires optimization potential
    \item Frequent EV charging transactions, early EV connections, and PV surplus
increase optimization potential
    \item Two clusters in EV Charging Behavior: self-optimizing and overnight charging
\end{highlights}


\end{frontmatter}




\section{Introduction}



Electrification of vehicles is the central strategy to decarbonize the transport sector \cite{IPCC.2022, IPCC.2023}. Accordingly, the EU targets at least 30 million emission-free vehicles on its roads by 2030, with the goal of nearly all vehicles being emission-free by 2050. As electric vehicles are mostly charged  in private homes, their electrification has huge implications for household energy systems and management. In Germany alone, authorities estimate that private households will install 5.4 to 8.7 million charging stations by 2030 \cite{Commission.2020, Arnhold.2020}, with nearly all expected to be coupled to rooftop photovoltaic (PV) systems, making prosumer households central to the energy transition by integrating renewable energy generation with transportation electrification. 

This raises new questions. Households face the challenge of balancing intermittent energy production with increasingly dynamic electricity demand profiles and growing electrical loads. 
In addition to the goal of reducing $CO_2$ emissions, balancing production and consumption becomes an economic optimization: increased self-consumption minimizes electricity bills. Home Energy Management (HEM) strategies are the answer to this optimization problem. HEM strategies have been developed to optimize energy flows in residential buildings, including the use of Energy Storage Systems (ESS) and shifting the  consumption of large flexible loads, such as residential EV charging (which alone is comparable to the average electricity usage of a single-person household \cite{MeinAuto.2023}). 

The common practice in HEM applications involves rule-based (RB) approaches that optimize energy flows using simple, predefined rules based on domain knowledge \cite{Fischer.2017}. However, these RB approaches offer limited flexibility. More sophisticated methods, such as Model Predictive Control (MPC), use mathematical optimization to derive optimal schedules for charging and deploying flexible loads. Despite their advantages, MPC methods largely depend on the quality of forecasts and the accuracy of the physical models \cite{Langer.2022}. As optimization challenges become increasingly complex and dynamic, and situations more difficult to predict, there is a growing need for more adaptable and flexible approaches. Model-free Deep Reinforcement Learning (DRL) has emerged as a promising method to address the dynamic and uncertain environment of HEM optimization problems. By applying state-of-the-art Artificial Intelligence (AI) methods, DRL enables an agent to observe its environment and learn individually adapted HEM strategies, offering a more flexible and robust solution. While a growing body of research explores applying DRL algorithms to address this challenge, many studies rely on simulated data  \cite{Han.2023}. Further research is needed to understand how DRL agents respond to real-world data when optimizing prosumer households with EV chargers.

The contribution of this work is first to identify the context and circumstances when algorithmic control in home energy management can realize relevant environmental and economic benefits. This is achieved, secondly and in contrast to other research by employing here a unique, novel dataset of full-year data with real-world measurements from nine different households on fixed-rate electricity tariffs located in the German-Austrian-Swiss  (DACH) region. It includes measurements for PV generation, battery storage (dis-)charging, household load, and real EV charger usage. This allows for the analysis of real-world impacts. Third, we compare different algorithmic controls of HEM and specifically train and test the DRL Deep Deterministic Policy Gradient (DDPG) model in a new architecture, including EV charging within a PV and Battery Energy Storage Systems (BESS) architecture under fixed prices, which emulates a typical prosumer context in the DACH region. We evaluate this new algorithmic control against two benchmarks: the RB and the MPC approach.

The paper is structured as follows. Chapter Two reviews the current literature on household EV charging behavior and optimization in HEM. 
{Chapter Three details the system architecture, including optimization task and Reinforcement Learning approach.} Chapter Four describes the experimental setup and explains the benchmarking. We report and interpret the results in Chapter Five, adding a simulation study with hyperparameter tuning on synthetic data. Chapter Six covers the discussion of implications and limitations, while Chapter Seven concludes.


\section{State-of-the-Art in prosumer household energy management}

\subsection{Optimization behavior of prosumer households}

Households equipped with PV systems, battery energy storage systems (BESS), and EVs leverage advanced energy management strategies to optimize their energy consumption and enhance self-sufficiency. These systems allow prosumers to align energy usage patterns with PV generation by utilizing intelligent scheduling tools and optimization algorithms. For instance, home energy management systems (HEMS) deploy machine learning and distributed control strategies to determine when to store energy in batteries or directly use it for appliances or EVs. Such optimization is typically most effective during periods of high solar irradiance, when surplus energy can be stored for later use or directed to flexible loads like EVs \cite{Chapman2018, Cui2023}. (\ref{apx:resEVcharging} provides details on residential EV charging).

Self-optimization in households relies heavily on effective energy scheduling to minimize reliance on grid electricity during peak hours. For example, EVs can be scheduled to charge during midday when solar generation peaks, and BESS  can discharge stored energy during evening peak periods to meet household demands. Additionally, households can participate in time-of-use (TOU) tariff structures to lower electricity costs by shifting energy consumption to off-peak periods. These strategies are facilitated by predictive algorithms that forecast energy generation and consumption patterns, enabling households to maximize self-consumption and reduce energy bills \cite{Rose2019, Nadeem2018}.

Despite these technological capabilities, households often encounter situations where self-optimization is not fully realized. Behavioral aspects play a significant role in these suboptimal outcomes. For instance, high transaction costs associated with understanding and implementing optimization strategies can deter households from using their systems effectively. Additionally, the need to charge EVs at specific times, such as immediately upon returning home, may conflict with the ideal scheduling for maximizing PV generation or reducing grid dependency. Similarly, household members may prioritize convenience over optimization, such as using appliances at peak times rather than waiting for off-peak periods \cite{Kilthau2023, Thomsen2019}.

Technical challenges further compound these issues. Overvoltage risks during periods of excess generation can lead to curtailment of PV output, reducing the benefits of self-optimization. Limited battery storage capacity and degradation over time also restrict the ability of households to store and use excess energy efficiently. In some cases, households may lack sufficient real-time and forecasted data to make informed optimization decisions, leading to reliance on grid power even when alternative options are available. These barriers highlight the importance of integrating both technical solutions and behavioral factors to enhance the effectiveness of prosumer energy systems \cite{Rahim2023, Shewale2022}.

\subsection{Optimization Approaches}

A variety of optimization approaches are available or actively discussed in the literature. These range from rule-based (RB) approaches \cite{Fischer.2017, Elkazaz.2020, Li.2021b, Baraskar.2024} to using forecasts and rolling horizon optimizations in model predictive control (MPC) \cite{Pean.2019, Elkazaz.2020, Fitzpatrick.2020, Yousefi.2021, Seal.2023, Srithapon.2023, Thorsteinsson.2023}, as well as user-centered recommendation systems \cite{Shuvo.2022, Zharova.2024a, Zharova.2024b}. The optimization techniques include deterministic methods like mixed-integer linear programming (MILP) \cite{Golshannavaz.2018, Langer.2020, Fitzpatrick.2020, Athanasiadis.2023, Buechler.2019, Huy.2023}, heuristic \cite{Huang.2015, Liu.2020} and meta-heuristic methods \cite{Mary.2014, Rasheed.2022, Kelepouris.2023}, as well as stochastic techniques such as machine learning (ML) and AI \cite{Zupancic.2020, Rocha.2021, Zheng.2021, Huy.2023, Clift.2023}. 

We review the optimization methods through the concepts of flexibility, efficiency, adaptability, and complexity. Flexibility describes how well the system adjusts to changing conditions. Efficiency means achieving results with minimal resources. Adaptability shows how well the system learns and improves over time. Complexity reflects the system's design intricacy, affecting development and reliability. These factors are crucial for ensuring the performance, scalability, and practicality of HEM solutions in real-world applications. 

Comprehensive reviews  highlight the evolution of energy management strategies, the integration of emergent technologies, and the challenges and opportunities in HEM \cite{Ajitha.2023, Bakare.2023, Han.2023}. 
Deterministic methods provide precise optimization solutions but often require detailed models and can be computationally intensive. In contrast, stochastic methods, including DRL, offer flexibility and adaptability to dynamic environments. Here, we provide a brief  overview of the main approaches.


\textit{Default Approach} involves basic, manual strategies for managing energy usage. This can include simple methods such as power-mode EV charging, where the EV is charged at full power whenever it is connected. Another example is the use of manual time schedules for flexibility assets, such as setting specific times for heating or cooling systems to operate. While straightforward and low in complexity, this approach offers limited optimization potential and does not adapt to changing conditions, thus lacking in flexibility and adaptability.

\textit{Rule-Based Optimization} relies on preprogrammed rules derived from domain knowledge. These rules dictate how and when different energy assets should be used \cite{Pinamonti.2020}. For instance, a rule might specify that the BESS should only charge during periods of PV surplus. While this method can improve efficiency compared to the default approach and add some level of adaptability, it offers limited flexibility and cannot adjust to unforeseen changes in energy production or consumption, thereby limiting its overall efficiency and adaptability.

\textit{Model Predictive Control} is a more advanced optimization technique that forecasts PV production and household energy usage. Based on these forecasts, MPC calculates and follows an optimal path for energy management, adjusting for deviations as they occur. This approach allows for more precise and dynamic optimization, ensuring that energy usage is aligned with PV generation and minimizing reliance on grid electricity. MPC enhances flexibility and adaptability by adjusting to forecast errors in real-time, making it particularly effective in scenarios where accurate forecasting is possible \cite{Langer.2022}. However, the complexity of this method can be high, requiring significant computational resources and is reliant on the availability of precise forecasts.

\textit{Deep Reinforcement Learning}  represents the cutting edge of optimization in HEM. Unlike RB or MPC approaches, DRL learns from past decisions and continuously improves its performance. It makes dynamic, informed choices based on real-time data, allowing for highly adaptable and efficient energy management. DRL can optimize complex systems with multiple interacting components, making it well-suited for modern prosumer homes with diverse energy assets \cite{Antonopoulos.2020, Ali.2023}. The adaptability and flexibility of DRL are unparalleled, as it can handle a wide range of scenarios and dynamically adjust to new patterns. However, this comes with increased complexity and a need for extensive computational power and sophisticated algorithms.


\subsection{Deep Reinforcement Learning in HEM}



Existing research on DLR in HEM is structured mainly around three main streams:  Optimization and Control, Human-Centric Approaches, and Evolutionary Algorithms and Heuristics. Each group offers a unique perspective and addresses specific challenges inherent to HEM.


Research on Optimization and Control investigates advanced DRL techniques to enhance efficiency and model performance. Studies into scalable multi-agent systems with DRL \cite{Wang.2024} and the use of safe DRL techniques for building management \cite{Wang.2025} demonstrate the potential of DRL to address complex energy management challenges.
Real-time demand response strategies \cite{Li.2020b} and DRL approaches paired with forecasting methods \cite{Ren.2022} underscore the adaptability of DRL frameworks to dynamic environments, improving energy management outcomes.


Human-Centric Approaches integrate human feedback and behavior into DRL models to optimize energy use while ensuring user comfort. Research on DRL-based home energy recommendation systems \cite{Shuvo.2022} and  models that integrate residents' activities \cite{Li.2020a} highlight the importance of user-centric designs in enhancing home energy management.


Lastly, the Evolutionary Algorithms and Heuristics stream explores hybrid approaches blending DRL with methods like particle swarm optimization and heuristic models. Heuristic-based real-time control models integrating DRL \cite{Rezaeimozafar.2024} are also explored for  managing residential PV-battery systems.

DRL has led to significant progress in the field, leading to the development of highly effective algorithms for decision-making in complex environments \citep{Zhang.2023, Sumiea.2024}.
\textit{Deep Deterministic Policy Gradient} (DDPG) is a prominent algorithm in DRL that 
holds great promise for HEM due to its ability to handle continuous action spaces and learn optimal policies in complex environments. 
Our focus is on  whether DDPG can offer additional advantages for HEM households by leveraging real-world data. 



Recent research on DRL with DDPG for optimizing residential HEM can be categorized into two main streams: Standard DDPG and Enhanced DDPG algorithms (see Table~\ref{tab:DDPGinHEM}). The Standard DDPG stream focuses on application of DDPG to optimize residential energy management. These works demonstrate the versatility of DDPG in integrating energy systems, balancing efficiency, and maintaining user comfort. For example, an energy management framework leveraging vehicle-to-grid and vehicle-to-home modes \citep{Yi.2021}. Similarly, DDPG is employed to minimize energy costs while ensuring comfortable indoor temperatures \citep{Yu.2020}. 
Efforts to integrate bidirectional EV charging and room temperature control with DDPG showcase significant energy savings and enhanced comfort \cite{Svetozarevic.2022}. 
Additionally, DDPG is used for integrated HVAC and BESS control, demonstrating cost savings and load flexibility in real-world testing \cite{Touzani.2021}. 
DDPG can further help to achieve high self-sufficiency without compromising comfort when optimizing smart home systems with heat pumps and PV  \citep{Langer.2022}.

\begin{table}
    \small
    \centering
    \caption{Articles using Deep Deterministic Policy Gradient for Home Energy Management. Abbreviations: Battery Energy Storage System (BESS); Energy Storage System (ESS); Heating, Ventilation and Air-Conditioning (HVAC); Real-Time-Pricing (RTP); Thermal Energy Storage (TES); Time-Of-Use (TOU). 
    }
    \resizebox{\textwidth}{!}{%
    \begin{tabular}{lccccccc}
    \toprule
         &  PV& ESS& HVAC&EV&  Pricing&  Dataset& Location\\
    \midrule
\multicolumn{8}{l}{\textit{Standard DDPG}}\\
         \cite{Yi.2021}&  x& BESS& -&bi-directional&  TOU&  real-world (one household) & USA (LA, CA)\\
\cite{Yu.2020}&  x& ESS& x&-&  TOU&  real-world (Pecan Street)& USA (Austin, TXS)\\
 \cite{Svetozarevic.2022}& -& -& x& bi-directional& TOU& real-world & CHE\\
  \cite{Touzani.2021}& x& ESS& HVAC&-& TOU& real-world experiment& USA (Berkeley, CA)\\
         \cite{Langer.2022}&  x& BESS \& TES& HVAC&-&  fixed&  simulated&  USA (Chicago) \& DEU\\

\multicolumn{8}{l}{\textit{Enhanced DDPG }}\\          \cite{Zenginis.2022}&  x& ESS& x&-&  RTP&  not explained
& -\\
         \cite{Huang.2022}&  x& BESS& -&-&  TOU&  simulated& USA\\
         \cite{Ye.2020}&  x& BESS& HVAC&-&  TOU&  not explained
& -\\
         \cite{Kodama.2021}&  x& BESS& HVAC&-&  RTP&  simulated& JPN (Tokyo)\\
         \cite{Li.2021}&  -& BESS& -&-&  RTP&  simulated& USA (NE)\\
  \cite{Zhang.2024}& x& ESS& -& x& RTP& real-world (loads simulated)&AUS\\ 
     \bottomrule
 This work& x& BESS& -&x& fixed& real-world (9 households) &DEU, AUT, CHE\\
    \end{tabular}
    }
    \label{tab:DDPGinHEM}
\end{table}


The Enhanced DDPG algorithms incorporate innovations designed to improve performance in residential settings. For instance,  clustering-based reinforcement learning is introduced to train multiple agents on homogenous data subsets, improving efficiency \cite{Zenginis.2022}. Hybrid approaches such as combining DDPG with deep Q-learning address discrete-continuous action spaces \cite{Huang.2022}, while prioritized experience replay enhances real-time autonomous control \cite{Ye.2020}. Multistep prediction methods further improve learning efficiency \cite{Kodama.2021}. Integration of DDPG into smart home platforms enables optimized task scheduling and energy management \cite{Li.2021}. Additionally, multi-agent algorithms, such as an interior-point policy optimization-enhanced DDPG, address uncertainties in secure energy management \cite{Zhang.2024}. 

While most studies focus on systems combining PV, BESS, and demand management, only few include EVs, emphasizing early-stage bidirectional charging technologies 
\citep{Yi.2021} \citep{Svetozarevic.2022}. Fixed-pricing structures, a predominant tariff model in regions such as  Germany, remain largely underexplored, with only a single study addressing this topic \citep{Langer.2022}. 

The data sources and deployment methods in these studies vary significantly. Real-world data, such as single-family house measurements \citep{Yi.2021} or the Pecan Street dataset from Austin, Texas \cite{Yu.2020}, are used in some instances. Real-life experimental deployments are seen rarely \citep{Svetozarevic.2022},
\citep{Touzani.2021}. However, most works rely on simulated data, using standard PV generation and load curves to validate optimization methods. While simulations are valuable for developing efficient methods, they often neglect household diversity and real-world variations, such as EV connection patterns. We address this gap by  investigating diverse prosumer behaviors and real-world data to better understand the practical implications of residential energy optimization.

\section{System Architecture}
\label{chap:theoreticaloptpot}

\subsection{Optimisation Task}

\begin{figure}
    \centering
    \includegraphics[width=0.7\linewidth]{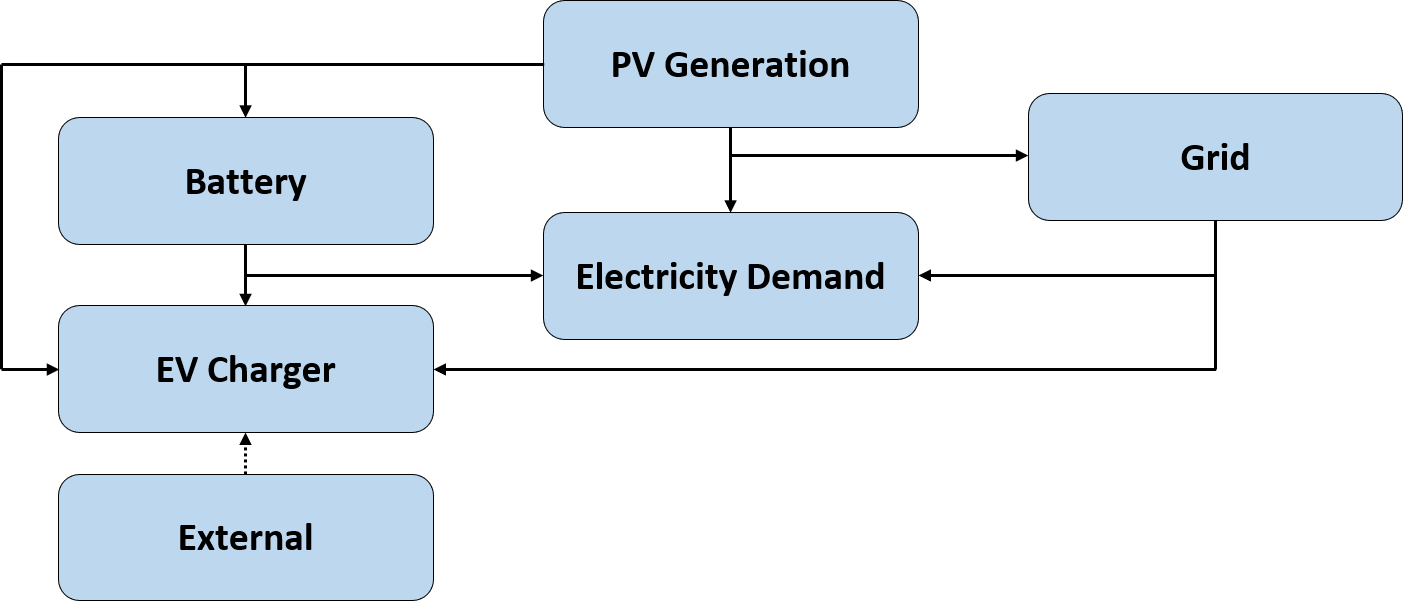}
    \caption{System architecture}
    \label{fig:DRL_Architecture}
\end{figure}

Our system architecture, inspired by \cite{Langer.2022}, represents a typical setting of a prosumer household (Figure \ref{fig:DRL_Architecture}). It consists of a PV system, a BESS, a residential EV charging station, and a connection to the electricity grid. The electrical demand of the household is summarized as Electricity Demand. The connection to the EV is not permanent, since it can either be connected (present) or disconnected (away). Each time window in which the EV is connected is called a charging transaction. To account for the special case that the EV charging demand is not fully met by the end of a charging transaction, the architecture includes the option of an External source, such as a public charging station.

The optimization goal is to minimize electricity purchase costs by optimally controlling the battery charging and discharging  ($b_t^c, b_t^d$) and EV charging  demand ($d_t^{ev}$), while ensuring the EV is fully charged (equation \ref{eq:minimization}). Drawing electricity from the grid  ($ X_t^{gr \to {}}$) or externally  ($ X_t^{ext \to {}}$) is priced at $p_{buy} $, and feeding electricity into the grid ($X_t^{ \to gr})$ is remunerated at feed-in price $ p_{sell}$.

\begin{equation}
\min_{b_t^c, b_t^d,  d_t^{ev}} \sum_{t} \left( p_{buy} \times (X_t^{gr \to demand} + X_t^{gr \to ev_{}} + X_t^{ext \to ev_{}}) - p_{sell} \times X_t^{pv \to gr} \right)
\label{eq:minimization}
\end{equation}

Under fixed pricing conditions with  $p_{buy} > p_{sell}$, minimizing costs aligns with maximizing PV self-consumption. PV surplus (i.e., the excess PV energy after meeting household demand) is therefore critical in the optimization of energy flows. The task is to determine optimal charging schedules for the BESS and the EV to maximize PV-surplus-use while ensuring the EV is fully charged to the required level during its connection period.

The default charging mode of EV chargers, referred to as power-mode, involves charging at full power from the moment the EV connects until it is either fully charged or disconnected. Optimizing for PV self-consumption involves delaying EV charging to periods when PV surplus is available. For EV charging transactions that commence during periods of sufficient PV surplus, power-mode charging is inherently optimal. This is also applicable to scenarios where no PV surplus is available throughout the entire connection period. In the latter, it makes no difference at what time the EV is charged. Moreover, the optimization involves storing any remaining PV surplus in the BESS for later use, which is subject to power limits and efficiency losses. Due to these inefficiencies, charging or discharging the BESS from or to the power grid (under fixed price condition) is economically non-viable.

The system architecture and constraints lead to the following prioritization of electricity flows:

\begin{enumerate}
    \item Any electricity generated by the PV system meets the current household electricity demand.
    \item Any PV surplus fulfills the EV charging demand and BESS charging demand according to the energy management strategy.
    \item Grid balancing:
    \begin{enumerate}
        \item Further PV surplus is fed into the grid.
        \item Residual demands are met by drawing electricity from the grid.
    \end{enumerate}

\end{enumerate}

\subsection{Reinforcement Learning}

\begin{figure}[H]
    \centering
    \includegraphics[width=0.4\linewidth]{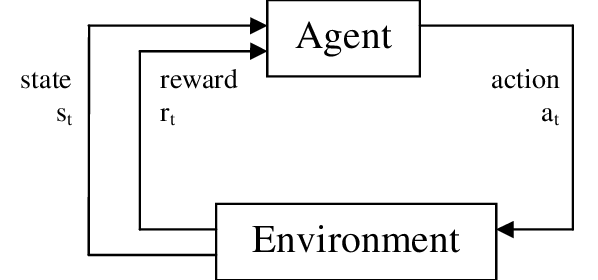}
    \caption{Markov Decision Process \citep{Sutton.1997}}
    \label{fig:MDP}
\end{figure}

RL focuses on making decisions by learning the optimal behavior in environments to maximize a reward signal. It is characterized by a Markov Decision Process (MDP) with states and actions (see Figure \ref{fig:MDP}).
The MDP is a mathematical model used to describe decision-making scenarios where outcomes are influenced by both randomness and the actions of a decision-maker. In the MDP, an agent makes decisions by selecting actions based on a strategy, called policy. The objective is to find an optimal policy that maximizes the expected cumulative reward over time (\citep{Howard.1960, Puterman.1994, Zhang.2023, Sumiea.2024}). The following components describe the MDP:

\begin{itemize}
    \setlength{\itemsep}{0pt}  
    \setlength{\parskip}{0pt}
    \item {State Space (S)} - the set of all possible states in which the system can exist
    \item {Action Space (A)} - the set of all possible actions that the agent can take
    \item {Transition} - the transitions from one state to another given a specific action
    \item {Rewards (r)} - the immediate rewards received after transitioning from one state to another due to an action
    \item {Value Function} - the expected cumulative reward of being in a state and following a particular policy.
\end{itemize}

In the following, we describe our system's architecture using the MDP framework.

\paragraph{States}

We adapt the representation of the states  (equation \ref{eq:states}) from \cite{Langer.2022} and include 8 features: the state of charge  (SoC) of the BESS ${soc}_t^b$, the SoC of the EV ${soc}_t^{ev}$, the countdown that counts the hours until the EV is disconnected ${c}_t^{ev}$, the temporal representation of the day-time via cosine $h_t^{\text {cos}}$ and sine $h_t^{\text {sin}}$, the season ${season}_t$, as well as the household electrical demand $d_t^h$ and the PV generation $g_t^{pv}$. The SoC of the EV is exogenous at the start of each charging transaction and endogenous during transactions. For all time steps $t$ where the EV is not connected the ${soc}_t^{ev} = 1$ and the countdown ${c}_t^{ev} = -1$. 

\begin{align}
\label{eq:states}
\mathbf{s}_{\mathbf{t}}=[\underbrace{soc_t^b, \overbrace{soc_t^{ev}}^{\text{exogenous}}}_{\text {endogenous}}, \underbrace{c_t^{ev}, h_t^{\text {cos }}, h_t^{\text {sin }}, season_t, d_t^h, g_t^{pv}}_{\text {exogenous }}] .\\
\    \hspace{2.5em} \underbrace{\phantom{\text{deterministicxxxxxxxxxxxxxxxxxx}}}_{\text {deterministic}}
    \underbrace{\phantom{\text{uncert}}}_{\text {uncertain }}
    \nonumber
\end{align}

\paragraph{Actions}

The continuous actions (equation \ref{eq:actions}) represent the optimization decisions for the charging control of the BESS and the EV charger. Learning target-SoCs instead of specific load amounts stabilizes the learning process \cite{Langer.2022}, therefore, we define the actions for the BESS and the EV as target-SoCs ${Tar}_t^b$ and $ {Tar}_t^{ev}$ represented by values between [0, 1].

\begin{equation}
\label{eq:actions}
\mathbf{a}_{\mathbf{t}}=[{Tar_t^b, Tar_t^{ev}}]
\end{equation}

\paragraph{Transition}

This work assumes a deterministic environment, meaning that the transition from one state to the next, given an action, occurs without uncertainty.
While connected, the EV charges ($d_t^{ev}$) according to the chosen target value ${Tar}_t^{ev}$. The system then either charges the BESS up to the target value using residual PV surplus ($b_t^c$) or discharges in order to meet any residual demand ($b_t^d$). All charging amounts are subject to power limits (${ev}_{rate}^{max}$ and ${b}_{rate}^{max}$). After EV and BESS charging, electricity exchanges with the grid balance out any further PV-surplus or further residual demand (grid purchase $x_t^p$ or feed-in $x_t^f$) incurring costs or earning a profit.
The input data does not include any excessive electrical household demands and all other generation and load variables have sensible power limits. Therefore, we omit a power limit for the grid connection.
The balancing constraint combines the three system inputs and four system outputs. The SoCs in $t+1$ are subsequently set according to the charging and discharging loads ($d_t^{ev}$, $b_t^c$, $b_t^d$). The ${soc}_t^b$ is adjusted for efficiency losses. 
Equation  \ref{eq:ev_charging} describes the EV charging. The BESS charging and discharging are given by equations \ref{eq:bess_charging} and \ref{eq:bess_discharging}, respectively. Finally, equation \ref{eq:system_balance} represents the overall system balance:

\begin{equation}
d_t^{ev}= \max[0, \min[\underbrace{ev_{rate}^{max}}_{\text {Inverter }}, \underbrace{({Tar}_t^{ev} \times soc_{max}^{ev}) - soc_t^{ev}}_{\text{Load to reach target SoC}}]] \times connected 
\label{eq:ev_charging}
\end{equation}

\begin{equation}
b_t^c = \min[\underbrace{(g_t^{pv}-d_t^h-d_t^{ev})}_{\text {Residual PV }}, \underbrace{b_{rate}^{max}}_{\text {Inverter }}, \underbrace{({Tar}_t^b \times soc_{max}^b) - (1 - loss^{b}) \times soc_t^b}_{\text{Load to reach target SoC}}]
\label{eq:bess_charging}
\end{equation}

\begin{equation}
b_t^d = \min[\underbrace{(d_t^h-d_t^{ev}-g_t^{pv})}_{\text {Residual demand }}, \underbrace{b_{rate}^{max}}_{\text {Inverter }}, \underbrace{{soc}_t^b}_{\text{Current SoC}}]
\label{eq:bess_discharging}
\end{equation}

\begin{equation}
\underbrace{g_t^{pv} + b_t^d + x_t^p}_{\text{System input}} = \underbrace{d_t^h + d_t^{ev} + b_t^c + x_t^f}_{\text{System output}}
\label{eq:system_balance}
\end{equation}

\paragraph{Reward}

The base reward observed by the agent after each step consists of any revenue or costs from exchanges with the grid. At the end of charging transactions, the reward is reduced by the real external and the virtual costs of not charging the EV as desired (equation \ref{eq:reward}). The real-external cost can be understood as the cost of having to finish charging the EV at a public charging station. The virtual cost represents the added cost of discomfort, modelled as a quadratic function $\text{discomfort\_weight} \times (1 - \text{soc}_t^{ev})^2$, similar to the approach used in \cite{Jin.2020} and \cite{Hao.2022}. Lastly, a linear penalty term further reduces the reward, penalizing any ${Tar}_t^{ev}$ actions below 1 while the EV is disconnected. This aids the learning of the agent, which consequently learns to focus on setting actions for  ${Tar}_t^{ev}$ on time periods when the EV is present.

\begin{align}
    \label{eq:reward}
    r_t = p_{sell} \times X_t^{pv \to gr} - p_{buy} \times (X_t^{gr \to demand} + X_t^{gr \to ev_{}} + X_t^{ext \to ev_{}}) \\
    -\ \text{discomfort\_weight} \times (1 - \text{soc}_t^{ev})^2 \times {disconnect}_t \nonumber \\
    - \ \text{penalty\_weight} \times (1 - \text{Tar}_t^{ev}) \times disconnected_t \nonumber
\end{align}

\paragraph{Value Function}

The RL agent observes the rewards and develops an optimal policy $\pi (a|s)$ for choosing actions given certain states. The optimal policy is the policy that maximizes the sum of discounted rewards. The value function in equation \ref{eq:valuefunction} estimates the Q-value $Q^*(s,a)$, the reward of the current period $r_t$ plus the future discounted rewards given a reward-maximizing policy $\pi$ (\citep{Zhang.2023, Sumiea.2024}).
\begin{equation}
\label{eq:valuefunction}
Q^*(\operatorname{s}, \operatorname{a}) = \mathbb{E} \left[ r_t + \gamma \ \max_{\operatorname{a}} Q^*(\operatorname{s}_{t+1}, \operatorname{a}_{t+1}) \mid \operatorname{s}_t = \operatorname{s}, \operatorname{a}_t = \operatorname{a} \right]
\end{equation}
\begin{equation}
    \pi^*(s) = \arg\max_a Q^*(s, a)
\end{equation}

To enhance performance in high-dimensional state spaces, RL integrates Deep Learning, where Deep Neural Networks (DNN) represent the agent’s decision-making policy. 
DDPG utilizes an actor-critic architecture, where the actor-network generates actions and the critic-network evaluates these actions by estimating the Q-value. Both networks are trained simultaneously to optimize the policy and value functions \citep{Silver.2014, Zhang.2023, Sumiea.2024}.
DDPG employs two key techniques for learning stability: the replay buffer and target networks. The replay buffer stores a diverse set of experiences, allowing the algorithm to sample from past interactions, which helps prevent overfitting and improves stability. The target networks, which include a target actor and a target critic, update more slowly using a soft-update technique. This approach mitigates the instability caused by the moving target problem in Q-learning. By integrating these techniques, DDPG enables robust learning in complex environments with continuous action spaces \citep{Sumiea.2024, Zhang.2023}. We provide the details on DDPG Implementation in \ref{ddpg_impl}.

To illustrate the application of DDPG in this study, consider the following scenario: the actor encounters a state characterized by favorable weather conditions, low household load, and the knowledge that the EV will remain connected for four more time steps. Subject to exploration noise in training, the actor may decide to charge the EV at full power and allocate the remaining surplus to charge the BESS for later use. The agent then observes the reward zero, as no PV surplus was fed into the grid to generate profit. Concurrently, it observes an increase in the SoC of both the EV and the BESS. With higher SoCs, subsequent time steps necessitate less charging for the EV to avoid penalties. Furthermore, the BESS can discharge, which lessens the dependency on purchasing electricity from the grid. Over time, the critic-network learns that higher SoCs ultimately lead to higher rewards in future time steps, essentially learning the value function. Guided by the critic’s evaluations, the actor will eventually select actions that maximize cumulative future rewards given a particular state.

\section{Experimental Setup}

\subsection{Data}
\label{chap:data}

The empirical study is based on real-world data obtained from a residential energy storage system provider. The measurements stem from the company’s battery management system of 90 households equipped with PV systems, BESS, and EV chargers located in the DACH region (Germany, Austria, and Switzerland). The data spans from November 2020 to October 2021. The numerical features consist of the date and time of each measurement,  power readings 
for the PV system generation, total household load, and EV charger load, aggregated at 15-minute intervals. The categorical features include an anonymized household ID (also referred to as ChargerID) and an anonymized transaction ID, which uniquely numbers each charging transaction within a household. 
Additionally, the data includes aggregated charging measurements for each EV charging transaction, specifying the start and end times and total energy charged. This enables detailed analysis of individual charging sessions, independent of the continuous time series data 
(we provide descriptive statistics in \ref{apx:data}).

Many households do not have continuous data for the entire period from November 2020 to the end of October 2021. Due to the importance of seasonality in PV generation and EV usage patterns, the DRL implementation focuses on nine households with nearly complete data (less than 5 hours of missings are addressed with quadratic interpolation). 
We use the data from 
all 90 households for a comprehensive user behavior analysis.

A resampling into hourly values lowers the computational complexity of the study, by reducing the number of time steps for each household to 8760. We split the data into 180 days (4320 hours) for training, 60 days (1440 hours) for evaluation, and 125 days (3000 hours) for testing. To preserve seasonal variation, we create segments of about 15 days for training, 5 days for evaluation, and 10 days for testing, always splitting at midnight, when PV generation is zero and household load is typically low. The algorithm adjusts the exact length of each interval, ensuring that overnight EV charging transactions remain intact.


\subsection{Benchmarking}

The lower benchmark is a Rule-Based approach for BESS charging combined with Power-Mode EV charging (RBPM). In  RBPM, the BESS charges to 100\% whenever PV surplus is available and discharges to 0\% whenever there is residual demand. The EV charges at full power to 100\% from the moment it is connected. RBPM represents the widely used energy management strategy in residential setups.

We use MPC with full information as the upper benchmark. Perfect forecasts are assumed, meaning the model receives PV production, household load, EV connection times, and loads for all time steps as inputs. Our implementation extends the MPC model from \cite{Langer.2020} by incorporating EV charging. The MPC optimization with a set of constraints matches the same conditions as in the DRL model and solves for maximum profit. This benchmark represents the theoretical optimum, i.e. an upper limit that is unattainable without full information. It is useful for evaluating how closely the DRL approach approximates this theoretical optimum. 
The difference between RBPM and DRL is a theoretical optimization potential. The objective of DRL is to surpass the lower benchmark, with its performance evaluated by the extent to which it realizes this optimization potential. 


Due to the sensitivity of DRL to randomness, we conduct training, evaluation, and testing for each of nine household using 40 different random seeds. 
The primary performance metric is the total daily profit derived from grid exchanges. To compare our DRL model to the benchmarks, we measure the realization of the optimization potential. Additionally, a discomfort score quantifies the average shortfall in meeting the charging demand per transaction. A score of 1 indicates an average shortfall of 1 percentage point in the SoC, meaning the EV was on average charged to 99\% instead of 100\%. We conduct evaluation runs to assess various settings and fine-tune hyperparameters, while we use testing data to produce the final performance results. For details on DRL experimental setup we refer to the \ref{apx:ddpg_setup}.



\section{Results}

\subsection{DRL performance depends on optimization potential}
\label{chap:modelresults}

We report the results as profit per day for the benchmarks and the DRL model and specify the realization of optimization potential (see Table \ref{tab:ResultsPreTest}). 
For the DRL, we report the mean  result on test data set across all 40  agents to evaluate the general performance of the model (the results on training and evaluation data sets are provided in \ref{apx:train_eval_results}). We further highlight the results of the respective agent that performed best in evaluation
, and the discomfort score. 

\begin{table}[h]
\centering
\caption{Performance evaluation results for nine households}
\resizebox{\textwidth}{!}{%
\begin{tabular}{c|ccc|ll|ll|ll}
\toprule
& \multicolumn{2}{c}{\textbf{Benchmarks}}&& \multicolumn{5}{c}{\textbf{DRL}} &\\
 & Lower& Upper& Optimization& \multicolumn{2}{c}{Result}& \multicolumn{2}{c}{Potential Realized}&\multicolumn{2}{c}{Discomfort}\\
 \textbf{Households}& (RBPM)& (MPC)& Potential& Mean& Best Eval& Mean& Best Eval&Mean&Best Eval\\
 
\midrule

 &\multicolumn{5}{c}{Mean profit from grid interaction [€/day]}& & & \multicolumn{2}{c}{Percentage-points}\\
\midrule
Household 01& 0.22& 0.49& 0.27& 0.30& 0.34& 32\%& 46\%&1.37&1.33\\
Household 02& -2.24& -2.23& 0.01& -2.31& -2.24& 0\%& 0\%&2.50&1.00\\
Household 03& -0.83& -0.76& 0.07& -1.01& -0.97& 0\%& 0\%&3.01&4.77\\
Household 04& -0.32& -0.12& 0.20& -0.42& -0.36& 0\%& 0\%&3.52&2.69\\
Household 05& -0.16& -0.05& 0.11& -0.22& -0.17& 0\%& 0\%&4.43&1.19\\
Household 06& -5.96& -5.87& 0.09& -6.09& -6.09& 0\%& 0\%&4.48&2.86\\
Household 07& -5.62& -5.61& 0.01& -5.66& -5.63& 0\%& 0\%&4.35&4.52\\
Household 08& -3.60& -3.56&0.04& -3.64& -3.58& 0\%& 38\%&4.39&3.68\\
Household 09& -4.17& -3.82& 0.36& -4.04& -3.96& 38\%& 58\%&1.32&3.40\\
\end{tabular}
}
\label{tab:ResultsPreTest}
\end{table}

The DRL model  beats the lower benchmark for households 01 and 09, though with a moderate realization of the optimization potential. It achieves 32\% and 38\% for the mean profit,  and 46\% and 58\% by the model agents that exhibit the best results. 
We record the highest difference in profit for the best-eval agent on dataset 09, which generates 0.21€/day more than the RBPM. The discomfort score remains below 4.5\% throughout.



We observe a significant relationship between optimization potential, i.e. the difference between MPC and RBPM results, and the performance of the DRL model. We structure the interpretation of results by dividing the households into three groups and incorporating findings from our descriptive analysis (see \ref{apx:data}):

\begin{enumerate}

    \item Households with {marginal to no} optimization potential: 02, 03, 07, and 08
    \item Households with {low} optimization potential: 04, 05, and 06
    \item Households with {moderately high} optimization potential: 01 and 09
\end{enumerate} 

\paragraph{\textbf{Marginal to no} optimization potential}

These households display the lowest optimization potential, ranging from 0.00 to 0.07€/day. In comparison to the other households, infrequent charging transactions and short charging durations characterize their EV charging. We observe limited PV surplus availability (especially in 02 and 07) and connection times that suggest manual optimization.

As discussed in Chapter \ref{chap:theoreticaloptpot}, power-mode EV charging is inherently optimal for transactions that either commence during periods of sufficient PV surplus availability or do not exhibit any PV surplus throughout. Additionally, the BESS compensates for inefficient power-mode charging according to capacity and charging power limits. The absence of substantial optimization potential implies that using the power-mode for EV charging, in combination with a BESS as a buffer, is efficient in this group.

The benchmark results also imply that a simple RB approach offers a nearly optimal solution for the BESS charging schedule. The rule involves charging when a PV surplus is available and discharging when there is an electricity deficit. MPC can only marginally improve over RB BESS-charging by maintaining the SoC at the lowest necessary level, which avoids standing losses (0.003\% of the SoC).

The results of the DRL agents with the best evaluation performance are comparable to the profits of the nearly optimal RBPM benchmark, falling short by 0.00 to 0.04€/day. This  indicates that the DRL agent 
learns to follow a policy that resembles the rules in the RB approach for BESS charging. Random exploration in the continuous action space leads to some imperfections, explaining minimally lower profits. We conclude that faced with a lack of optimization potential in this group, the DRL agent approximates the efficient RBPM approach.

\paragraph{\textbf{Low} optimization potential}

The second group of households exhibit a low potential for improvements over the lower benchmark, ranging mostly from 0.09 to 0.20€ profit per day. 
For this group, the descriptive analysis (\ref{apx:data}) shows moderate to high EV connection frequency with varying daytimes. Household 05 connects the EV for predominantly short durations. Household 06 is characterized by low PV surplus availability combined with the largest BESS. This explains a limited amount of optimization potential stemming from EV connections. Comparing RBPM to the optimal solution indicates the presence of some situations in which power-mode EV charging is inefficient.

While the DRL model surpasses the lower benchmark in some  instances on evaluation data, even the best-evaluated model runs fail to generate more profit than RBPM on test data. Notably, the profit gap to the lower benchmark is larger for these households compared to the group without any substantial optimization potential. This suggests that the frequency of optimizable EV charging transactions is insufficient for the DRL model to learn a consistently efficient EV charging policy. Consequently, it does not perform well when applied to test data, especially where most connection times are ideal for power-mode EV charging.

\paragraph{\textbf{Moderately High} optimization potential}

Two households reliably exhibit the highest optimization potential across training, evaluation, and testing, ranging from 0.27 to 0.51€/day. Our descriptive analysis (\ref{apx:data}) reveals relatively high EV connection times, particularly for household 09, coupled with frequent PV surplus availability, notably in houselhold 01. Despite occasional charging patterns suggesting manual "self-optimization" by users, these households display a higher incidence of inefficient (optimizable) EV connection times compared to the others. Both households employ the smallest and least powerful BESSs (6.75kWh usable capacity, 3.3kW inverter power).

In this group, the DRL model exhibits superior performance, surpassing the lower benchmark not only with the best-evaluated agents but also in average  results across agents. This outcome underscores that, given sufficient EV connection times and optimization potential, the DRL agents effectively learn and implement policies that yield higher profits compared to power-mode charging.



\subsection{Frequent EV charging transactions, early EV connections, and PV surplus increase optimization potential}

To better understand the conditions when the DRL model has more potential to save costs for households, we create a synthetic dataset. 
In our simulation study, household 06, which offers a balance of EV charging frequency and durations, serves as a base to derive a synthetic dataset. We double the EV charging transactions and shift the connection time to start before typical PV surplus times. To address the shortages of PV surplus, we increase PV production by a factor of 1.5. To limit the impact of the BESS as a buffer for inefficient EV charging, our synthetic household includes the smallest BESS model (6.75kWh capacity and 3.3kW inverter power). We train, evaluate, and test in the same way as for the households 01 to 09. We further use the synthetic data to fine-tune the hyperparameters in the DRL model (see \ref{chap:simul} for details). 

As expected, the changes to the input data lead to a much higher optimization potential at 1.13€/day in the synthetic dataset (Table \ref{tab:ResultsSynt}), more than triple the optimization potential of any of the original datasets. Given this input, the tuned DRL model can on average realize over 70\% of the optimization potential, 76\% when using the agent with the best evaluation results. The results of the DRL model on synthetic data confirm our findings that the DRL performance relies on sufficient optimization potential in the input data.
We further apply the fine-tuned model with reduced actor and critic network sizes on our real-world dataset of nine households. The results show slight improvements of 0.01 to 0.07€/day in the results for most households (Table \ref{tab:ResultsFinTest} in Appendix).

\begin{table}[h]
\centering
\caption{Performance evaluation results for the synthetic household on test data}
\resizebox{\textwidth}{!}{%
\begin{tabular}{c|ccc|ll|ll|ll}
\toprule
& \multicolumn{2}{c}{\textbf{Benchmarks}}&& \multicolumn{5}{c}{\textbf{DRL}} &\\
 & Lower& Upper& Optimization& \multicolumn{2}{c}{Result}& \multicolumn{2}{c}{Potential Realized}&\multicolumn{2}{c}{Discomfort}\\
 \textbf{Household}& (RBPM)& (MPC)& Potential& Mean& Best Eval& Mean& Best Eval&Mean&Best Eval\\
 
\midrule

 &\multicolumn{5}{c}{Mean profit from grid interaction [€/day]}&&&Percentage-points&\\
\midrule
 Synthetic & -4.09& -2.96& 1.13& -3.28& -3.22& 71\%& 76\%&4.81&2.88\\
\end{tabular}
}
\label{tab:ResultsSynt}
\end{table}


\subsection{DRL agent effectively aligns  charging of  EV and  BESS with  PV surplus}

\begin{figure}
    \centering

 \begin{subfigure}[b]{\textwidth}
        \centering
        \includegraphics[width=0.9\textwidth]{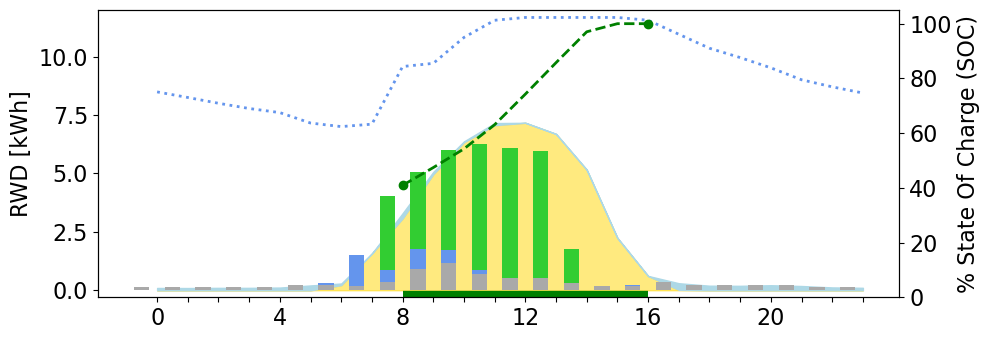} 
    \end{subfigure}

     \begin{subfigure}[b]{\textwidth}
        \centering
        \includegraphics[width=0.9\textwidth]{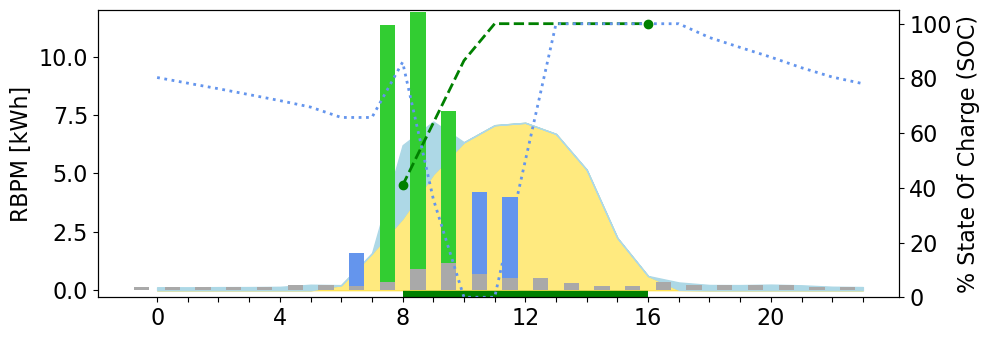} 
    \end{subfigure}

     \begin{subfigure}[b]{\textwidth}
        \centering
        \includegraphics[width=0.9\textwidth]{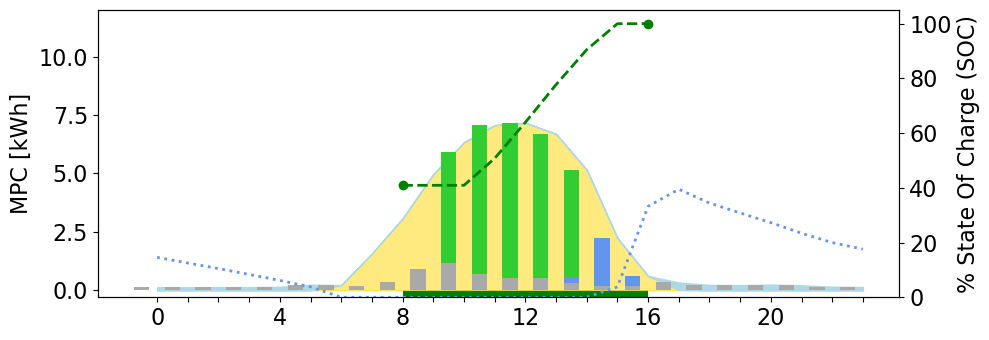} 
    \end{subfigure}

     \begin{subfigure}[b]{\textwidth}
        \centering
        \includegraphics[width=0.9\textwidth]{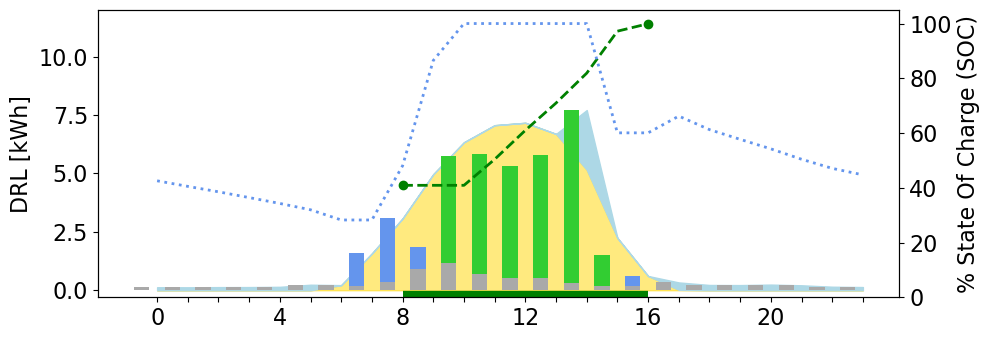} 
    \end{subfigure}

     \begin{subfigure}[b]{\textwidth}
        \centering
        \includegraphics[width=0.7\textwidth]{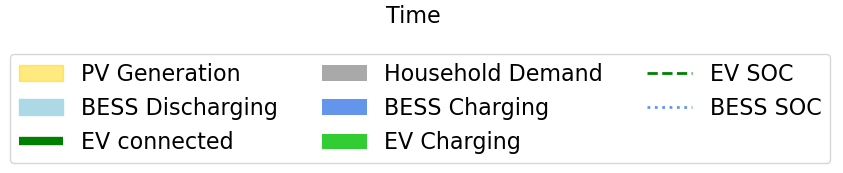} 
    \end{subfigure}
   \caption{
    Household 01 on March 08 with real-world data, RWD (panel 1), RBPM (panel 2), MPC (panel 3), and DRL (panel 4).  The stacked curves with filled-in areas represent available electricity (PV generation and BESS discharging), while the stacked bars represent electrical demands (household demand, EV charging, and BESS charging).   
   }
    \label{fig:RB_MPC_DRL01}
\end{figure}

To demonstrate the performance of the DRL agent in practice, we select one day (8th of March) in the household 01 data and compare the results to the benchmarks (Figure \ref{fig:RB_MPC_DRL01}). 
Household demand, PV generation, and EV connection times remain consistent across four cases. In all cases the BESS capacity is sufficient to meet the household demand in the morning and at night of that day. The first panel in the figure shows the real-world behavior of the household.


The RBPM  (panel 2) uses the initial PV surplus generated between 6:00 and 8:00 entirely for charging the BESS. Once the EV is connected, it charges at full power for 2 to 3 hours, exceeding the current PV generation by 8.27kWh at 8:00, 6.94kWh at 9:00, and 1.33kWh at 10:00. The BESS discharges, subject to the inverter limit, to compensate for 3.13kWh and 2.25kWh of the deficit in the respective hours. Grid purchases meet the remaining deficit totaling 11.16kWh over three hours. RBPM fully charges the EV by 11:00, the BESS in the following two hours, and feeds any further PV surplus into the grid. To understand the impact on total profit, consider the following calculation: 
Given the fixed price of 0.41€/kWh for grid purchase and 0.09€/kWh for feed-in, shifting grid consumption of 11.16kWh to a time of PV surplus would have avoided costs of 3.57€ on that day.

The MPC approach  under full information (panel 3) demonstrates an optimal solution to the problem. It perfectly aligns all electricity usage with PV generation, eliminating the need for grid purchases altogether. The BESS only charges up to approximately 40\%, which suffices to meet all electrical demand until the next PV surplus period on the following day. By charging the BESS as minimally and as late as possible, this approach minimizes the efficiency losses associated with charging and storing electricity. 

The DRL agent (panel 4) exhibits a clear improvement over the lower benchmark. It delays and then distributes the EV charging over the time window from 10:00 to 16:00. PV generation fully covers demands in all but one hour: From 2:00 to 3:00, the BESS discharges approximately 1.5kWh into the EV. Despite not having access to future  PV generation and household consumption data, as in the MPC approach, the DRL agent effectively aligns the charging of the EV and the BESS with the PV surplus. Thereby it avoids the need to draw electricity from the grid. However, the match is not perfect, and the BESS capacity is fully utilized despite its inefficiencies.

\subsection{EV Charging Behavior: self-optimizing and overnight charging}
\label{chap:EV_behav}


We examine EV charging transactions across  90 households, focusing on the timing and the energy of transactions. To ensure the data represents active EV users,  transactions with connection durations exceeding 48 hours (2.07\%
) are excluded. Additionally, we remove recordings shorter than 30 minutes to filter out  false and test connections (10.70\%). 
The remaining 8,471 transactions have a median duration of 4.52 hours and a median charged energy of 10.75kWh. 
This indicates a high number of short charging transactions (below the median), with a steadily decreasing frequency for durations ranging from one to 8 hours.
For longer transactions, the distribution reveals a notable period between 12 to 18 hours where transactions occur more frequently compared to the 8 to 12-hours range (for details refer to  Table \ref{tab:ChargingTransactions} and Figure \ref{fig:combined_violin} in Appendix).

In three-quarters of the transactions, the charged energy is below 20kWh. For the remaining quarter, users charge their EVs up to 40kWh, with some outliers reaching nearly 85kWh. Given that typical EV battery capacities range between 20 to 120kWh \citep{UsableBatteryCapacity.2024},
we conclude that residential EV charging in this dataset is primarily characterized by smaller charging sessions, with relatively few cases of charging from a low battery level.
Notably, half of the transactions involve charging amounts below 10.75kWh, which can typically be achieved within one hour of charging at full power.  

To better understand user-specific patterns in charger usage across 90 households, we apply a k-means clustering approach using three features:  mean start time, mean end time, and mean duration of charging for each user. Using the elbow method to analyze the within-cluster sum of squares, we find that splitting households into two clusters is appropriate. Furthermore, the average silhouette scores, with a peak score of 0.440, indicate a moderate level of cluster separation for two clusters (see Figure \ref{fig:elbow} in Appendix for details).


The cluster analysis reveals clear patterns in charging behavior. The households in the first cluster, consisting of 45 users, typically begin charging transactions during the day and unplug in the morning, with mean start and end times of 13:00 and 10:00, respectively (Figure \ref{fig:TransactionTimesHistogram}). The average charging duration for this group, which we label as overnight chargers, is nearly 14 hours. The second cluster, also comprising 45 users, demonstrates significantly shorter charging transactions that generally start earlier in the day and end later in the afternoon. 
This behavior suggests a form of self-optimization, where users deliberately connect their EVs during periods of PV energy production (Figures \ref{fig:cluster1} and \ref{fig:cluster2} in Appendix provide details).



    



\begin{figure}[H]
    \centering
    \includegraphics[width=0.9\linewidth]{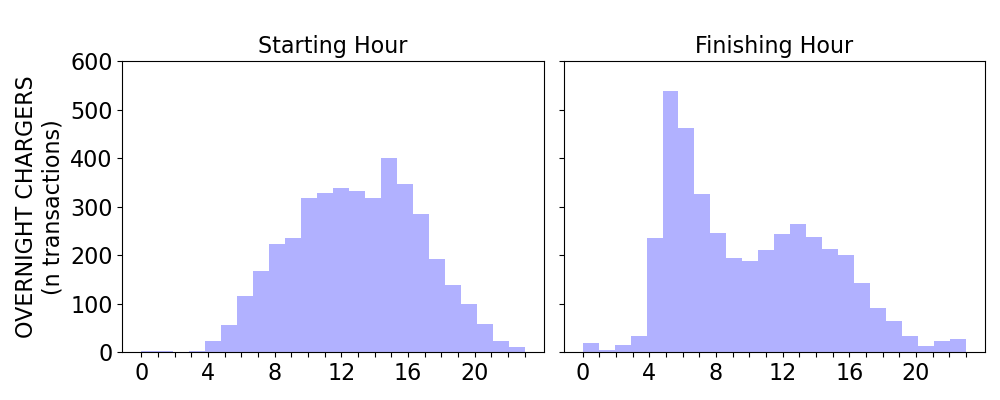}
    \vspace{-0.75cm}
    
    \includegraphics[width=0.9\linewidth]{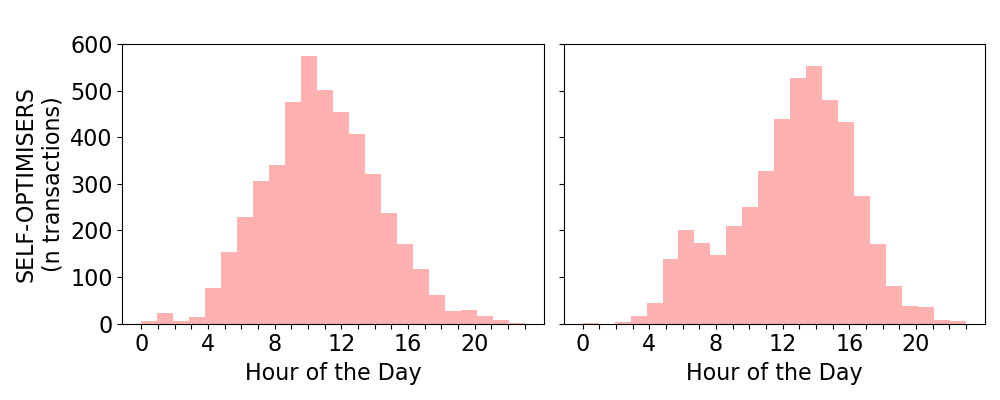} 
    \vspace{-0.25cm}


    \caption{Start and finish times of EV charging transactions for cluster of overnight chargers (panel 1) and cluster of self-optimizers (panel 2)}    
    \label{fig:TransactionTimesHistogram}
\end{figure}

 
\subsubsection{{Availability of PV-surplus and timing of EV charging transactions in the selected households explain the DRL results} }
\label{chap:9hous}

\begin{figure}[H]
    \centering
    \includegraphics[width=0.9\linewidth]
    {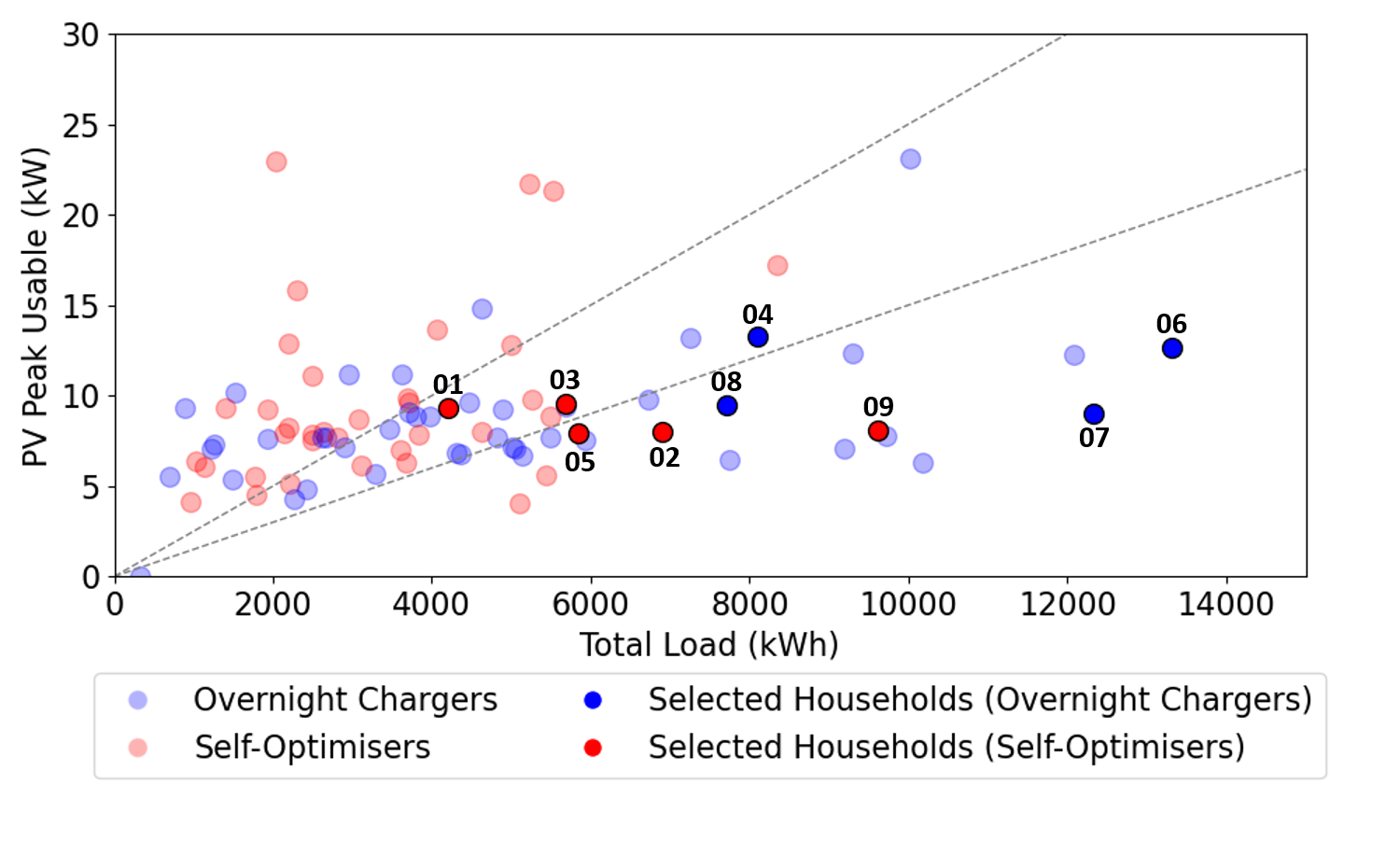}
    \caption{PV peak relative to yearly total household demand. The gray dashed lines indicate industry guidelines for PV system sizing, with factors of 1.5 and 2.5. The dots represent 90 households, with dark-colored dots highlighting the selected nine households.}
    \label{fig:PVload}
\end{figure}

To better understand the data used for the DRL model, we provide a detailed analysis of the selected nine households (see Figure \ref{fig:PVload}).  
Households 01, 02, 03, 05, and 09 belong to the cluster of ‘self-optimizers’, with moderate silhouette scores ranging from 0.450 to 0.643. Households 04, 06, 07, and 08 fit the category of ‘overnight chargers’, with silhouette scores ranging from low (0.274 for household 08) to moderate (for households 04, 06, and 07). Both clusters share the characteristic that most charging transactions start during daytime hours. However, despite belonging to one of the clusters, start and end times, as well as charging durations show significant variability, motivating the following in-depth analysis.


Optimization critically depends not just on connection times but also on the availability of PV surplus. The scatter plot in Figure \ref{fig:PVload} compares the PV production capacities to the annual electricity load of all 90 households, highlighting households 01 to 09. Industry guidelines for PV system sizing recommend multiplying the yearly electricity demand in MWh by a factor ranging from 1.5 to 2.5 to estimate suitable kW-PV-peak-sizes \citep{RegionalPhotovoltaik.2024, energieexperten.2024}. 
We derive a “usable” PV peak from the maximum PV production recorded in 15-minute intervals (\ref{apx:data}). The six households (02, 05, 06, 07, 08, 09) demonstrate PV peak usable sizes below the lower recommendation and three households (01, 03, 04) below the higher threshold, indicating that the availability of PV surplus for optimization may be too low overall. 


We summarize the total values for key variables across households 01 to  09, along with the mean totals for all datasets (Tables \ref{tab:descriptiveAll} and \ref{tab:ChargingTransactions} in Appendix). Households 01 and 04 show the highest ratio of PV production to the combined household and EV demand, with PV production roughly doubling the combined total demand. In contrast, household 02 has the lowest PV production overall. Households 06 and 07 have the highest household energy demands, almost double the average. These households are potentially equipped with a heat pump or similar large loads. Notably, the total PV production of household 07 is less than its total household demand, indicating a deficit in energy production.

\begin{figure}
    \centering
    \includegraphics[width=0.95\linewidth]{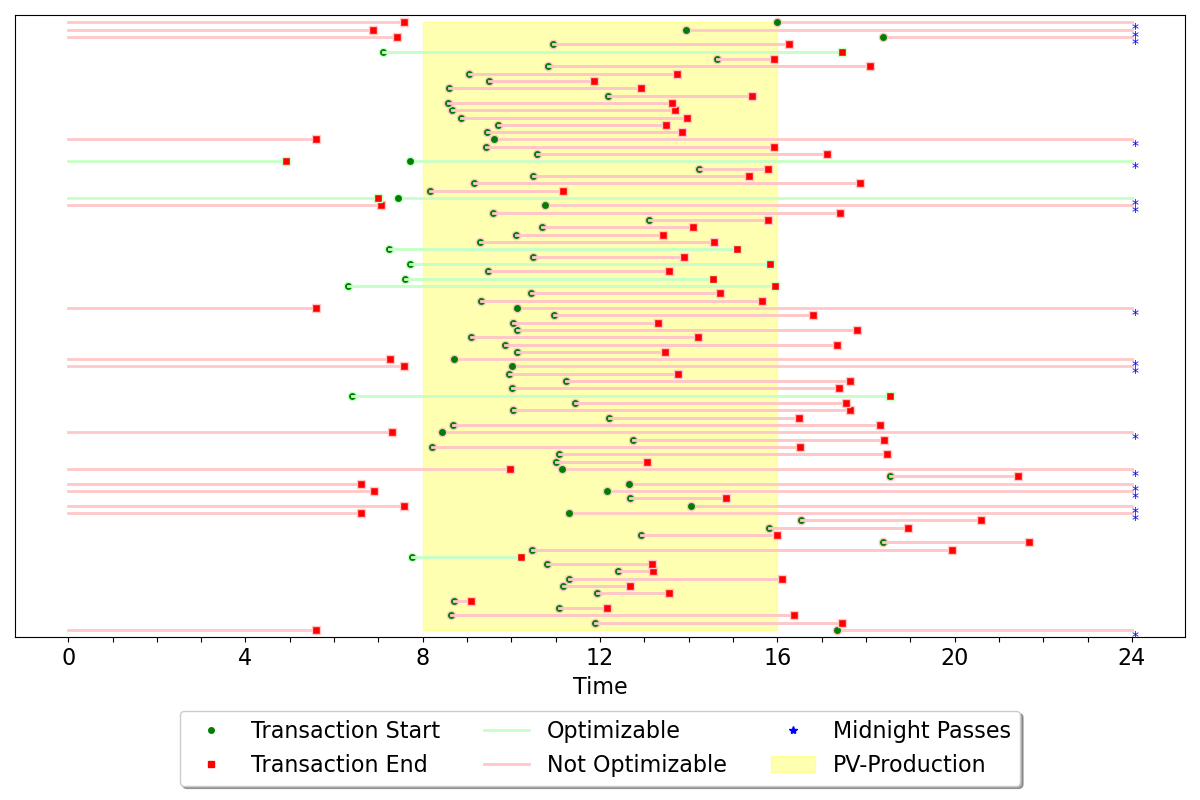}
    \caption{Timing of transactions and optimization potential (Household 01)}
    \label{fig:LineChart}
\end{figure}

Finally, in Figure \ref{fig:LineChart} we employ a line chart for a closer examination of each transaction and to assess the optimization potential for household 01 (see \ref{fig:LineChart2} to \ref{fig:LineChart9} in Appendix for other households). The chart displays all transactions as lines, spanning from their start time to their end time, while highlighting typical PV surplus hours from 8:00 to 16:00. Charging transactions that either commence during PV surplus hours or remain outside this time frame for their entire duration exhibit minimal optimization potential. Power-mode charging is already the optimal strategy in these cases that are consequently labeled as ‘Not Optimizable’ (red lines). Conversely, all other transactions are marked as ‘Optimizable Transactions’ (green lines). It is important to note that this classification provides an indication based on EV charging times and is not an exact measure. True PV surplus hours vary with the PV installation, season, and household demand. Based on the visual analysis of individual charging transactions, households 01, 04, and 09 exhibit the highest number of 'optimizable' transactions.


The descriptive analysis reveals that, overall, EV charging transactions within the population are typically short in duration and involve low energy consumption. Two distinct patterns emerge, short daytime charging, as well as long overnight charging transactions. It becomes apparent that some users conduct some form of manual optimization strategy by adjusting EV connection times to hours with sunshine. This behavior is accompanied by limited availability of PV-surplus. Notably, the timing of EV transactions and the availability of PV-surplus are crucial for optimization. This analysis suggests that the optimization potential across all nine households might be limited. This is particularly evident for households 02 and 03, which have particularly infrequent charging transactions, and households 02, 06, 07, and 09, which show relatively low PV surplus availability. Despite this, household 09 stands out with the highest number of charging transactions. In summary, the analysis implies that households 01 and 09 have the highest optimization potential.


\subsection{Grid Savings Potential}

\begin{figure}[h]
    \centering
    \includegraphics[width=0.8\linewidth]{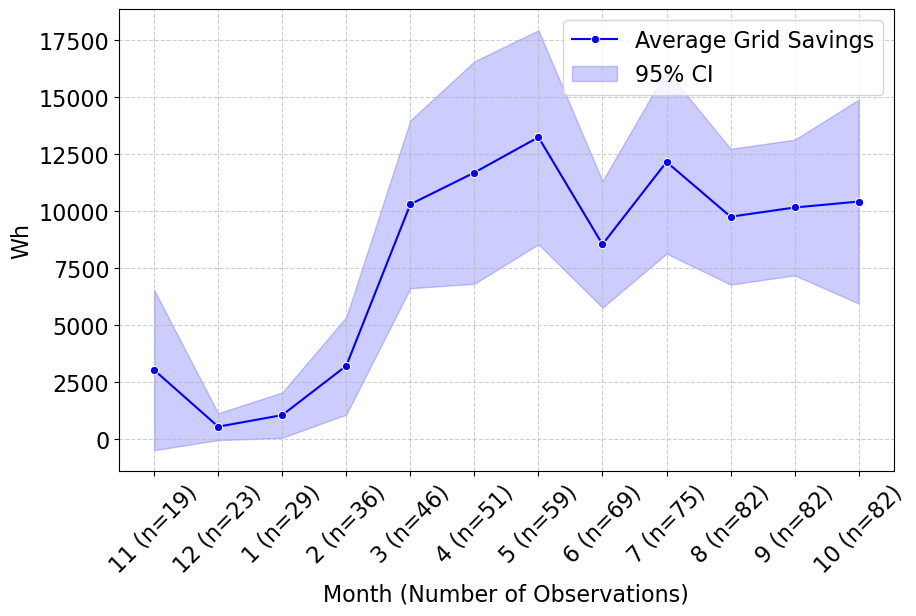}
    \caption{Monthly average potential grid savings from EV charging shifts for households with sufficient data. The potential savings represent the amount of grid usage during each EV charging transaction that can be offset by PV production if charging is shifted within the EV's connected period.}
    \label{fig:monthlysavings}
\end{figure}

We estimate the grid savings potential from EV charging shifts for 86 out of 90 households with sufficient data (Figure \ref{fig:monthlysavings}). 
Households are added throughout the year, and most do not contribute data for all 12 months. Additionally, some households exhibit sparse EV transactions. For each EV charging transaction, the potential grid savings are calculated as the minimum of the grid purchase for EV charging and the grid feed-in during the connected time. This reflects the idea that grid purchases could be reduced by shifting EV charging to periods when PV-surplus is available. Given the fixed grid price of 0.41€/kWh, the potential average annual cost savings per household amount to 37.65€   (Table \ref{tab:gridsavings} in Appendix). In terms of potentially avoided CO$_2$ emissions from energy production in Germany (with 0.45 kg CO$_2$/kWh), this would amount to 42.35 kg. 

On average, 2.6\% of grid consumption per household could be saved annually through optimal load shifts within the existing (partly optimized) charging schedule. However, the potential savings could be higher if EV charging were not optimized at all (i.e., in power mode). For comparison, when basing the calculation on simulated rule-based (basic/manual) EV charging, about 3.3\% of grid consumption per household could be saved annually. While grid savings are low in winter due to reduced PV generation, households could reduce grid usage by about 8.5 to 13.2 kWh per month per household on average between March and October. Furthermore, the household with the highest savings in a given month achieved monthly savings 5 to 15 times higher than the average household savings for that month. Thus, we assume that the potential for savings is significantly higher in households that charge their EV frequently and do not (yet) use smart charging strategies.




\hspace{1cm}

Overall, we conclude that our DRL model can learn an efficient policy for the optimization of EV and BESS charging. However, the random exploration in a continuous action space leads to imperfect policies. A simple rule-based approach is more suitable for optimal BESS charging when maximizing PV self-consumption in the context of fixed pricing conditions. With regards to EV charging, sufficient optimization potential is needed for the agent to learn an efficient policy. However, infrequent and short EV charging transactions, manual optimization behavior by users, and limited PV surplus availability, all combined with large BESS capacities as a buffer for inefficiencies lead to a lack of optimization potential.

\section{Discussion}

The MPC under full information perform better than DRL. However, this holds only in scenarios when accurate forecasting is possible. We argue, that in dynamic situations with high uncertainty the DRL optimization is a more suitable approach.


The implementation of the DRL model with DDPG showcases its capability to learn behavior and develop an EV charging strategy that optimizes the HEMS without requiring explicit rules or forecasts. The analysis of real-world data shows the DRL approach is effective in 2 out of 9 contexts, highlighting certain limitations inherent to the DRL method in specific real-life scenarios.

The potential for the DRL agent to outperform rule-based BESS and power-mode EV charging is constrained by several factors. Firstly, BESS charging in the context of PV-surplus optimization under fixed prices is easily represented by a simple rule, leaving little room for improvement through advanced algorithms. Secondly, short EV charging durations and predominantly daytime charging sessions, influenced by user behavior and self-optimization efforts, further limit the optimization potential. Notably, the measurement period from November 2020 to the end of October 2021 coincides with the COVID-19 pandemic. This period is marked by reduced mobility and an increase in home-office practices \citep{Wang.2022} likely influencing EV usage and charging behavior. Additionally, in scenarios lacking PV surplus the scope for optimization is minimal. To conclude, the effectiveness of the DRL optimization is highly dependent on sufficient training situations that include optimizable behavior, meaning the agent requires a rich dataset with diverse and representative scenarios to learn effectively. At the same time, a simple rule-based approach is nearly optimal for PV-surplus BESS charging aimed at maximizing PV self-use under fixed prices. This indicates that for straightforward scenarios, advanced algorithms may not provide substantial additional benefits over simpler methods.

The results also reveal that the DRL approach excels in more complex and dynamic situations. For instance, in scenarios involving longer EV charging transactions with sufficient PV that is inefficiently allocated by power-mode charging, the DRL agent demonstrates significant utility. Such situations, not easily representable by simple rules, benefit greatly from the adaptive learning capabilities of the DRL algorithm. Moreover, the results suggest that DRL is likely suitable for other complex and dynamic contexts, such as those involving dynamic pricing or multiple assets with dynamically varying efficiencies. This versatility indicates that the DRL approach has broader applicability in diverse energy management scenarios.

Lastly, the grid search for hyperparameter tuning yields only marginal improvements, with all but one of the parameters retaining the original value. This shows that the parameter choices identified by \cite{Langer.2022}, and proven effective in prior studies \cite{Fujimoto.2018, Lillicrap.2019, Raffin.2021, Yu.2021}, are also effective in this study's context. The consistency indicates that these parameters can yield reliable outcomes across diverse settings.

\subsection{Limitations}

This study employs technical simplifications in its representation of system components. Firstly, the efficiency of the battery inverter and EV charger is assumed to be fixed, which is inaccurate. In reality, power charging efficiency is dynamic with higher utilization generally resulting in lower efficiency losses. Secondly, the EV charging rate is considered fully flexible, whereas in practice, the speed of charging and the flexibility of the charging rate depend on the specific vehicle. Thirdly, battery degradation is disregarded although battery health is crucial in determining long-term cost and sustainability.

Regarding the system architecture, we use some assumptions to address the lack of specific data. For instance, EV SoC and capacity measurements are inferred from energy measurements. However, the ability to conclude the EV capacity from charged energy is limited. The EVs are likely never charged from empty, concealing the full capacity. Additionally, multiple different EVs could have been used per household. While knowing the true EV capacity is not crucial to our approach, it could add to the analysis of user charging behavior. 



Finally, the hyperparameter tuning for the DDPG part presents significant challenges. The tuning process is computationally expensive due to the number of potential parameters (over 15) that could be adjusted. Each setting has to be run multiple times with different random seeds to account for sensitivity to randomness. 
More efficient parameter tuning strategies 
could potentially enhance model results.

\section{Conclusion}

This work applies the Reinforcement Learning algorithm with Deep Deterministic Policy Gradient to a Home Energy Management System encompassing PV, BESS, and EV charging under fixed pricing conditions. 
Our findings indicate that while the DRL model effectively learns and optimizes EV charging strategies, its performance is context-dependent. Short daytime charging characterizes some households' EV connection times, suggesting users manually optimize by adjusting EV connection times with power-mode charging. Combined with frequent PV surplus shortages, this renders default power-mode EV charging optimal in most charging periods. Additionally, for BESS control aimed at maximizing PV self-consumption, a simple rule-based approach prevails. In scenarios with sufficient optimization potential, the model excels, learning behaviors that approximate rule-based BESS charging and optimally charge EVs. Our simulation study with synthetic data corroborates that DRL exhibits superior performance in datasets with greater optimization potential. Moreover, hyperparameter tuning confirms effective parameter choices that yield reliable outcomes across similar contexts. These insights underline the potential of DRL in managing intricate energy systems and dynamic environments, suggesting pathways for future research. 




\section*{Acknowledgment}

Alona Zharova greatly appreciates the support of the Joachim Herz Stiftung.

\appendix

\section{Residential EV Charging}
\label{apx:resEVcharging}

Recent years have witnessed significant technological advancements in residential EV charging. Residential EV chargers, otherwise referred to as Wallboxes, can be categorized into two types based on their power capacity: 1-phase chargers with a power capacity of 4.6kW, and 3-phase chargers capable of delivering up to 22kW. Notably, chargers with a power capacity exceeding 11kW require a permit \citep{Rudschies.2024}. Certain Wallbox manufacturers, particularly those operating in the HEM solutions sector, offer features called smart charging or PV surplus charging. These EV chargers integrate some form of optimized charging or usability enhancements and necessitate an online connection for operation \citep{go-e.2023, Spangenberg.2022}.

A promising technology for HEM is bi-directional charging which enable vehicle-to-grid (V2G) and vehicle-to-home (V2H) functions. These capabilities allow EVs to discharge stored energy into the grid or power homes \citep{Vollmuth.2024}. While the newest EV models enable this technology, it is still in its early stages. Widespread adoption will require further advancements and supportive legislation to facilitate broader application and integration into existing energy systems \citep{TheMobilityHouseTeam.2024}. For this reason, our work focuses on unidirectional charging.

Battery capacities of EVs currently available in the market vary significantly depending on the specific model and typically range from around 20 to 120 kilowatt-hours \citep{UsableBatteryCapacity.2024}. A salient challenge for utilizing EV chargers for load management is the necessity to read the EV's current SoC. The DIN ISO norm 15118-2 delineates the communication protocol between EVs and chargers and establishes standards for EV chargers to read SoCs \citep{ISO.2024}. However, the technical application of this standard is predominantly still under development for residential applications. Alternative solutions exist, such as aggregators that combine APIs with different EV manufacturers to enable services like smart charging, which includes the communication between the EV and the charger. Notable examples of such aggregators include Jedlix\footnote{https://www.jedlix.com} and Smartcar \footnote{https://smartcar.com}. Many applications on the other hand rely on manual user input to receive the SoC, which we also use as an assumption for our work.

\section{DDPG Implementation}
\label{ddpg_impl}

\subsection{DDPG Procedure}

\begin{figure}[H]
    \centering
    \includegraphics[width=1\linewidth]{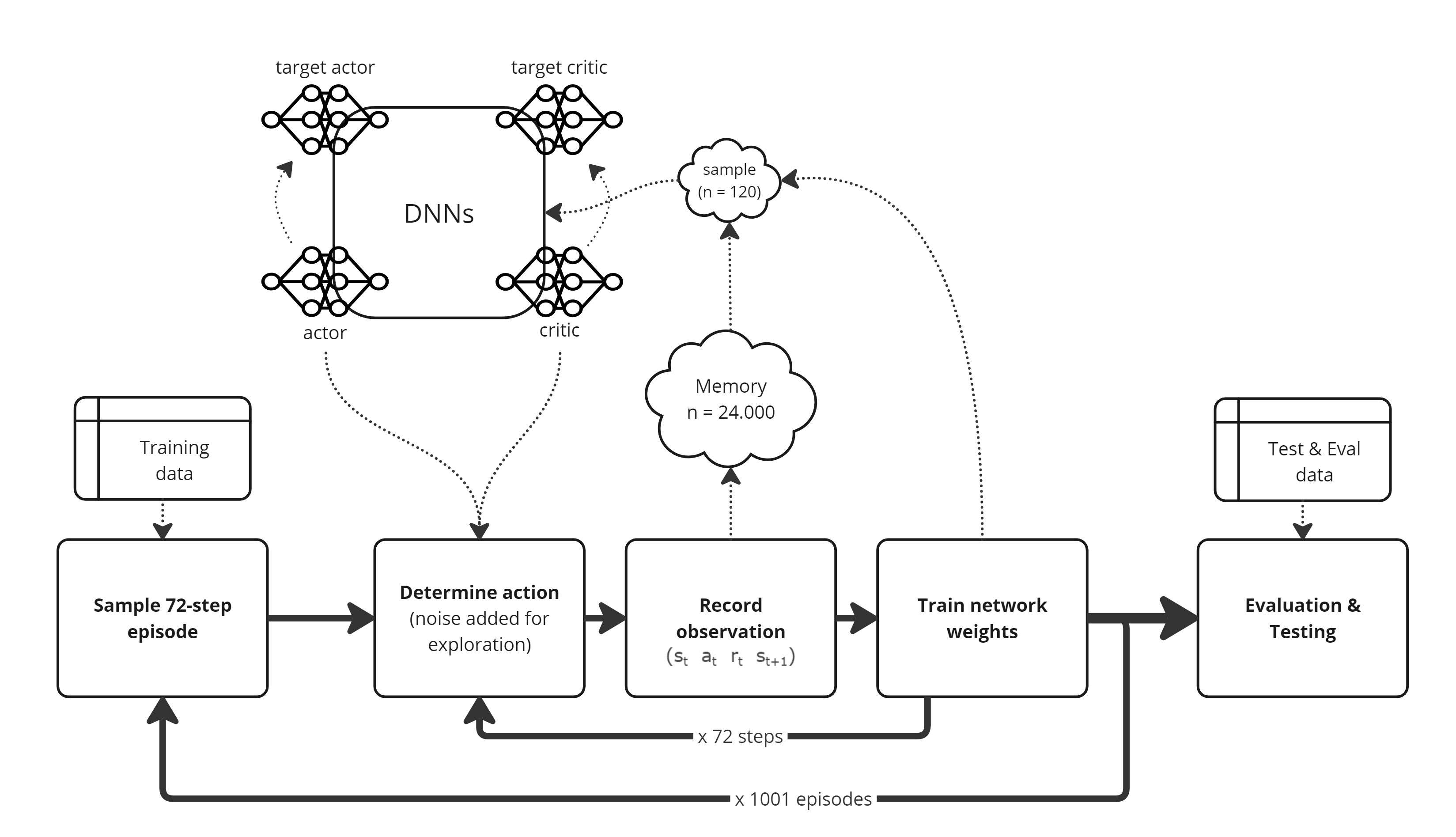}
    \caption{DDPG training procedure 
    }
    \label{fig:training_process}
\end{figure}

Figure\ref{fig:training_process} illustrates the DDPG procedure. In the model training, we apply the same method as for initializing the replay buffer to sample $M = 1001$ training episodes, each consisting of $H = 72$ hourly steps. For each step in each episode, the algorithm performs inference and updates network weights, as follows:

\paragraph{Inference}
 The actor-network selects an action based on the current state. Noise is added to the action to ensure exploration of the action space. The environment determines the transition to the next state $s_{t+1}$ and the reward $r_t$. The agent observes the result and stores a new tuple $[s_t, a_t, r_t, s_{t+1}]$ in the replay buffer, replacing the oldest entry.

\begin{figure}[H]
    \centering
    \includegraphics[width=1.0\linewidth]{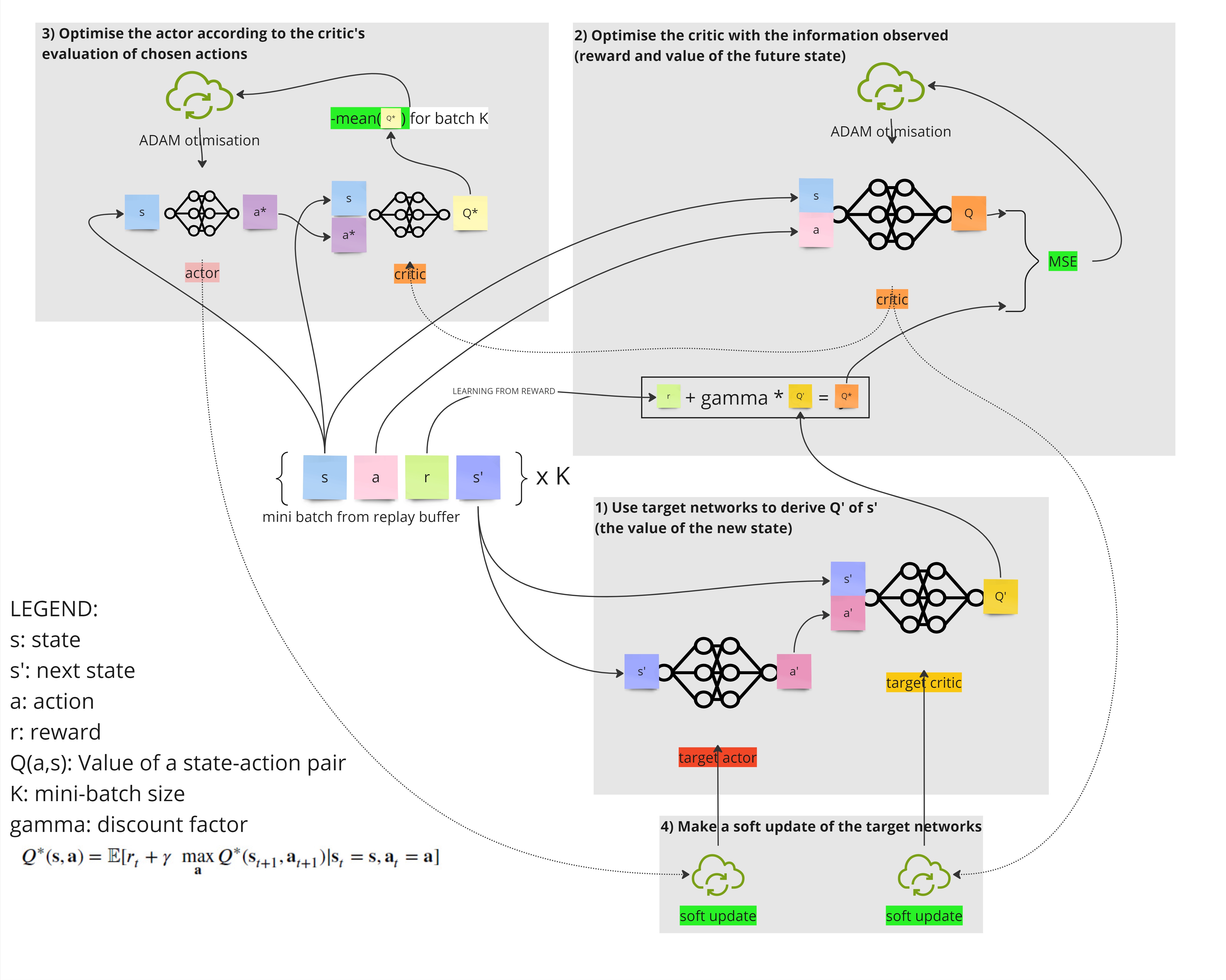}
    \caption{DDPG learning in each step 
    }
    \label{fig:DDPG_learn}
\end{figure}

\paragraph{Learning}
 The DDPG algorithm samples a random mini-batch of size $K = 120$ from the replay buffer and uses this batch to update the network weights. In Figure \ref{fig:DDPG_learn} we visualize the learning using a flow-chart with 4 stages:
 
\begin{enumerate}
    \item For each tuple \(i\) in the mini-batch \(K\), the target actor-network determines a hypothetical action \(a'\) (i.e., \(a_{t+1}\)) based on the tuple's "next state" \(s'\) (i.e., \(s_{t+1}\)). The target critic-network then uses \(s'\) and \(a'\) to estimate the expected Q-value \(Q'\) (i.e., \(Q^*(a_{t+1}, s_{t+1})\)) of the new state. This \(Q'\) value reflects the estimated value of the "next-state" \(s'\) in the tuple.
    \item Discounted by $\gamma$, \(Q'\) is added to the observed reward \(r\) to calculate the target Q-value \(Q^*\)  (i.e., \(Q^*(a_{t}, s_{t})\)) for the current time step \(t\). This \(Q^*\) represents the value of taking action \(a\) in state \(s\), incorporating immediate and estimated future rewards. The critic-network weights update using this information: The ADAM optimizer minimizes the mean squared error loss between \(Q^*\) and the critic’s estimated Q-value for each tuple in the batch.
    \item Stage 3 evaluates the actor-network using the updated critic-network: The ADAM optimizer adjusts the actor-network’s weights to minimize the negative mean Q-value over all Q-values determined by the critic-network from the tuples in batch \(K\).
    \item Finally, the target-networks perform a soft update at rate \(\tau\) using the actor and critic-networks respective weights. 
\end{enumerate}

\paragraph{Evaluation and Testing}
\label{chap:evaluationandtesting}
During the evaluation and testing phases for each household, the agents of 40 model-runs with different random seeds sequentially process the entire dataset, taking actions as determined by the respective actor-network. In contrast to the inference during training, no noise is added to the actions. The primary performance metric is the total daily profit derived from grid exchanges. To compare our model to the benchmarks, we measure the realization of the optimization potential. Additionally, a discomfort score quantifies the average shortfall in meeting the charging demand per transaction. A score of 1 indicates an average shortfall of 1 percentage point in the SoC, meaning the EV was on average charged to 99\% instead of 100\%. We conduct evaluation runs to assess various settings and fine-tune hyperparameters, while we use testing data to produce the final performance results.

\subsection{Programming and Model Setup}
\label{apx:ddpg_setup}
All codes and results are publicly available on GitHub\footnote{GitHub repository: https://github.com/Humboldt-WI/DeepRL\_4\_EV\_battery \_optimisation}. The code for the DRL model adapts from \cite{Langer.2022}\footnote{GitHub repository: https://github.com/lilanger/RL-SHEMS.git} in the programming language Julia (version 1.6.1)  \citep{Bezansos.2017}. It mainly uses the machine learning library Flux \citep{Innes.2018} and RL package Reinforce \citep{Breloff.2021}. The model-runs utilize CPU and GPU, as managed by the CUDA package \citep{Besard.2019}. We use Python (version 3.12.2) \citep{Python.2024} for the descriptive analysis, in the data selection and pre-processing as well as for the programming of the MPC Benchmark. The linear programming model utilizes the pulp package \citep{pulp.2024} in combination with the Gurobi solver under an academic license \citep{gurobi.2024}, while the conception draws from Julia code\footnote{GitHub repository: https://github.com/lilanger/SHEMS.git} by \cite{Langer.2020}.

Due to the sensitivity of DRL to randomness, we conduct training, evaluation, and testing for each dataset using 40 different random seeds. These are specific codes that control the sequence of random numbers generated for the exploration, ensuring reproducibility. For details on the construction of our random seeds, see \cite{Langer.2022}). A Bourne again shell (Bash) scheduler script executes the DRL code for 360 training and evaluation runs (40 seeds for nine datasets) on a Linux remote server by the Humboldt Lab for Empirical and Quantitative Research  \citep{LEQR.2024}, which is equipped with 192 GB of memory, 40 physical cores at 3.1 GHz, and two NVIDIA GPUs. The total runtime for training and evaluation across the nine datasets amounts to approximately 16 hours, depending on server utilization through other users. We provide the Bash scheduler script in GitHub$^3$.

While adopting most of the parameter choices from \cite{Langer.2022}, we triple the episode length (H) to 72, since this extended duration can accommodate 99.71\% of the charging transactions in the input data. A reduction in the number of training episodes (M) to one-third of the original count counterbalances the increased runtime, arriving at the same amount of individual training steps. The chosen discomfort weight of 0.01, implies that a 10\% deviation from the charging target reduces the reward by 1€ in the quadratic discomfort cost function in equation \ref{eq:reward}. Regarding the penalty weight, 0.1 proves to be effective in channeling the actor’s EV-SoC-related learning on the EV-connected times. We report all model specifications in Table \ref{tab:params}.

\begin{table}[H]
\centering
\caption{Parameter Settings of the DDPG Algorithm}
\begin{tabular}{cc|cc}
\toprule
\textbf{Name and symbol} & \textbf{Value} & \textbf{Name and symbol} & \textbf{Value} \\
\midrule
Data size Train/eval/test & 180;60;125 & Discount rate $\gamma$ & 0.99 \\
Number of episodes M & 1001& Episode length [hours] H & 72\\
Batch size K & 120 & Buffer size N & 24,000\\
Learning rate actor $\alpha_a$ & 0.0001 & Learning rate critic $\alpha_c$ & 0.001 \\
Soft update rate $\tau$ & 0.001 & Neural network size L & 300;600 \\
Exploration noise $\mathscr{N}$& $\mathscr{N}(0, 0.1)$& Optimizer & Adam \\
 Random seeds& 40& &\\
 Discomfort weight& 0.01& Penalty weight&0.1\\
\end{tabular}
\label{tab:params}
\end{table}

\subsection{Initialization}

The actor-network takes $[0,1]$ normalized values of the 8 states ($s_t$) as input and derives the action values ($a_t$) of range $[-1, 1]$ via a $tanh$ activation function. The critic-network outputs a Q-value for the state-action-pair when given the states and the actions as input. Both networks initialize with random weights, while the respective target-networks launch as copies.

Running inference involves applying actions to states and observing the rewards ($r_t$) and the transition to new states ($s_{t+1}$). To fill the initial replay buffer the algorithm repeatedly runs inference on randomly sampled 72h-episodes applying random actions at each hourly step. The replay buffer stores $N = 24,000$ observations as tuples $[s_t, a_t, r_t, s_{t+1}]$ and subsequently normalizes all states and actions using the entire set of observations.

The completion of EV charging transactions in this implementation is critical for learning Q-values because potential discomfort penalties for unmet charging demands are always triggered at the end of a transaction. For this reason, randomly chosen episodes that terminate during an incomplete charging transaction are either time-shifted to include the end or resampled if shifting is not possible. 

Episodes may also commence during an ongoing EV charging transaction. Consequently, it is essential to have valid EV SoC readings not only at the initiation of charging transactions but also throughout their duration. To address this, our implementation linearly interpolates the EV SoC for all time steps within each charging transaction in the training data. This ensures that episodes beginning amid an ongoing transaction can still be effectively utilized for initialization and training.

\subsection{Benchmark Results on Training and Evaluation Data}
\label{apx:train_eval_results}

\subsubsection{Results on Training Data}
\label{apx: train_results}

We detail the benchmark results and the optimization potential on the DRL training data, comprising 180 days, as opposed to 120 days for testing and 60 for evaluation in Table \ref{tab:ResultsBenchTrain}. The daily optimization potential spans between 0.02 to 0.51€/day, being highest for dataset 09, medium for 01 and 04, moderately low for 05 and 06, and lowest for 02, 03, 07, and 08. The last group exhibits 0.02 to 0.08€/day to optimize.

\begin{table}[H]
\centering
\caption{Benchmark Results on Training Data (180 days)}
\begin{tabular}{c|ccc}
\toprule
& &\textbf{Benchmarks}& \\
 & Lower& Upper& Optimization\\
 \textbf{HouseholdID}& (RBPM)& (MPC)& Potential \\
\midrule
HouseholdID 01& 1.29& 1.67& 0.38
 \\
HouseholdID 02& -2.40& -2.33& 0.07
 \\
HouseholdID 03& 0.16& 0.18& 0.02
 \\
HouseholdID 04& 0.13& 0.47& 0.34
 \\
HouseholdID 05& 0.24& 0.36& 0.12
 \\
HouseholdID 06& -7.34& -7.19& 0.15
 \\
 HouseholdID 07& -7.16& -7.12& 0.05
 \\
 HouseholdID 08& -3.31& -3.24&0.07
 \\
 HouseholdID 09& -4.34& -3.83& 0.51
 \\
\end{tabular}
\label{tab:ResultsBenchTrain}
\end{table}

\subsubsection{Results on Evaluation Data}
\label{apx:eval_results}

We report the evaluation results as profit per day for the benchmarks and the DRL model and specify the realization of optimization potential in Table \ref{tab:ResultsPreEval}. For DRL, we detail the mean and the best results across all 40 runs alongside the discomfort score (as defined in \ref{chap:evaluationandtesting}).

\begin{table}[H]
\centering
\caption{Evaluation Results (60 Days)}
\resizebox{\textwidth}{!}{%
\begin{tabular}{c|ccc|ll|ll|ll}
\toprule
& \multicolumn{2}{c}{\textbf{Benchmarks}}&& \multicolumn{5}{c}{\textbf{DRL}} &\\
 & Lower& Upper& Optimization& \multicolumn{2}{c}{Result}& \multicolumn{2}{c}{Potential Realized}&\multicolumn{2}{c}{Discomfort Score}\\
 \textbf{HouseholdID}& (RBPM)& (MPC)& Potential& Mean& Best Eval& Mean& Best Eval&Mean &Best Eval\\
 
\midrule

 &\multicolumn{5}{c}{Mean profit from grid interaction [€/day]}&&&\multicolumn{2}{c}{Percentage-points}\\
\midrule
HouseholdID 01& 0.12& 0.48& 0.36& 0.25& 0.38& 35\%& 72\%&2.24 &0.22\\
HouseholdID 02& -2.37& -2.36& 0.01& -2.44& -2.37& 0\%& 0\%&2.19 &2.05\\
HouseholdID 03& -1.25& -1.24& 0.01& -1.32& -1.26& 0\%& 0\%&7.11 &11.91\\
HouseholdID 04& -1.20& -1.02& 0.18& -1.30& -1.15& 0\%& 25\%&3.35 &1.66\\
HouseholdID 05& -0.27& -0.17& 0.10& -0.41& -0.32& 0\%& 0\%&2.09 &0.62\\
HouseholdID 06& -6.09& -5.81& 0.29& -6.03& -5.90& 23\%& 67\%&5.95 &7.64\\
 HouseholdID 07& -5.16& -5.16& 0.00& -5.21& -5.16& 0\%& 0\%&5.70 &4.28\\
 HouseholdID 08& -3.92& -3.91&0.02& -3.98& -3.92& 0\%& 0\%&4.39 &2.15\\
 HouseholdID 09& -4.64& -4.22& 0.42& -4.44& -4.32& 49\%& 77\%&1.37 &3.31\\
\end{tabular}
}
\label{tab:ResultsPreEval}
\end{table}

In the evaluation of datasets HouseholdID 01, 04, 06, and 09, the DRL model can beat the lower benchmark with the best-out-of-40 evaluation runs. These datasets carry the highest optimization potential. While the model realizes 25\% of the optimization potential in dataset 04, it reaches 67-77\% in 01, 06, and 09. Regarding mean DRL results, our model still outperforms the RBPM approach with 28-49\% realization in datasets 01, 06, and 09, while the RBPM generates a 0.10€/day higher profit in dataset 04. For HouseholdID 06, the discomfort per transaction of 5.95 percentage points on average and 7.64 for the best evaluation implies a noticeable discomfort for EV drivers. For the other three datasets on which DRL performed well, the discomfort per transaction is acceptably low, with less than a 3.5 percentage point deviation from the charging goal.

For the remaining 5 datasets, the DRL evaluation results are below or equal to the results of the lower benchmark. Notably, HouseholdIDs 02, 03, 07, and 08 exhibit minimal optimization potential: between 0.00 and 0.02€/day. This means that the RBPM performs almost or just as well as the perfect solution of MPC, leaving very limited potential for improvement. In these four cases, the best evaluation runs of DRL slightly deviate from MPC by 0.01 to 0.02€/day, with the mean performance trailing behind by 0.05 to 0.07€/day. Similarly to HouseholdID 06, the agent fails to charge the EV over 95\% on average for HouseholdIDs 03 and 07, causing a noticeable discomfort to the user. The last HouseholdID, number 05, displays a moderate to low amount of optimization potential (0.10€/day). Here, the model consistently produces lower profits compared to RBPM, 0.14€/day on average and 0.05€/day in the best run.

\section{Data and descriptive analysis}
\label{apx:data}
\label{chap:descriptiveanalysis}


\begin{table}[H]
\centering
\caption{Descriptive statistics of nine selected households}
\resizebox{\textwidth}{!}{%
\begin{tabular}{rrrrrr}
\toprule
\textbf{Dataset} & \textbf{Total PV }& \textbf{Total household}& \textbf{Total EV}& \textbf{EV charging}& \textbf{EV total time}\\
 & \textbf{production}& \textbf{demand}& \textbf{demand }& \textbf{transactions}&\textbf{connected }\\
 & \textbf{[kWh]}& \textbf{[kWh]}& \textbf{[kWh]}& \textbf{[count]}&\textbf{[h]}\\
\midrule
Household 01
&   10,147&2,746& 1,464& 73&641\\
Household 02
&  7,353&5,984& 924& 42&164\\
Household 03
&  9,598&5,233& 444& 43&180\\
Household 04
&  15,580&6,224& 1,872& 68&1020\\
Household 05
&  10,839&4,626& 1,214& 76&331\\
Household 06
&  15,946&11,926& 1,380& 68&1623\\
Household 07
&  9,718&11,695& 627& 28&297\\
Household 08
&  10,169&6,220& 1,487& 63&423\\
 Household 09& 9,808& 6,395& 3,217& 359&1841\\
\bottomrule
 Mean& 11,018& 6,783& 1,403& 91&724\\
\end{tabular}
}
\label{tab:descriptiveAll}
\end{table}

\begin{table}[H]
\centering
\caption{Statistics - All EV Charging Transactions}
\begin{tabular}{lll}
\toprule
\textbf{Metric}&  \textbf{Duration }&\textbf{Energy [kWh]}\\
mean&  8.45&13.97\\
std&  8.67&11.48\\
min&  0.50&0.00\\
25\%&  2.10&5.76\\
50\%&  4.52&10.75\\
75\%&  13.70&19.27\\
max&  47.89&84.34\\
\end{tabular}
\label{tab:ChargingTransactions}
\end{table}

\begin{figure}[h]
    \centering
    \begin{subfigure}{0.8\linewidth}
        \centering
        \includegraphics[width=\linewidth]{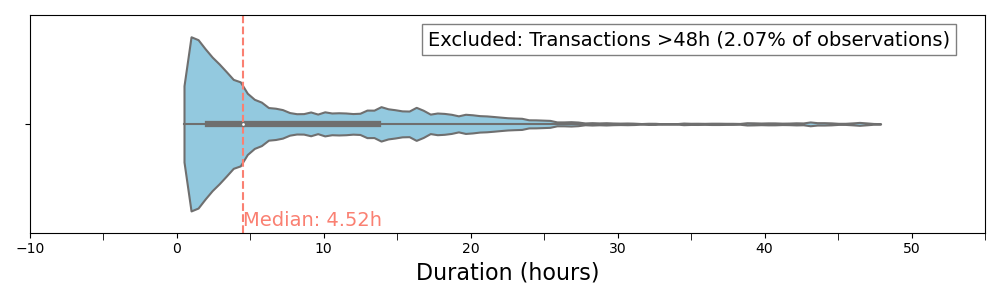}
        \label{fig:violinduration}
    \end{subfigure}
    \hfill
    \begin{subfigure}{0.8\linewidth}
        \centering
        \includegraphics[width=\linewidth]{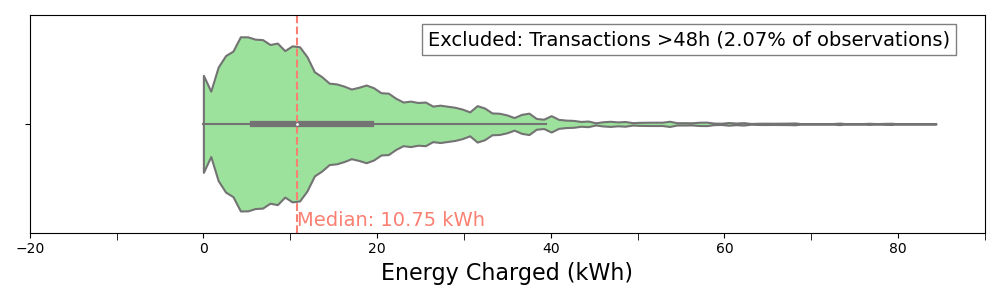}
        \label{fig:violinenergy}
    \end{subfigure}
    \caption{Violin plots for durations (h) and charged energy (kWh) of all EV charging transactions}
    \label{fig:combined_violin}
\end{figure}



\begin{table}[h]
\centering
\caption{Technical Setup of Households with EV chargers}
\resizebox{\textwidth}{!}{%
\begin{tabular}{lrrrr}
\toprule
\textbf{Dataset}&  \textbf{BESS capacity}&\textbf{Inverter power}& \textbf{EV capacity}&\textbf{PV peak usable}\\
 & $soc_{\text{max}}^{\text{b}}$& $b_{\text{rate}}^{\text{max}}$& $soc_{\text{max}}^{\text{ev}}$&$g_{max}^{pv}$\\
\midrule
Household 01&   6.75kWh&3.3kW& 48.25kWh&9.30kW\\
Household 02&  9.00kWh&3.3kW& 36.27kWh&8.02kW\\
Household 03&  9.00kWh&3.3kW& 45.51kWh&9.55kW\\
Household 04&  9.90kWh&4.6kW& 78.99kWh&13.28kW\\
Household 05&  9.00kWh&4.6kW& 37.21kWh&7.91kW\\
Household 06&  13.50kWh&4.6kW& 35.82kWh&12.63kW\\
Household 07&  10.80kWh&3.3kW& 36.52kWh&8.98kW\\
Household 08&  9.00kWh&3.3kW& 45.28kWh&9.46kW\\
 Household 09& 6.75kWh& 3.3kW& 21.94kWh&8.09kW\\
\end{tabular}
}
\label{tab:technicalsetup}
\end{table}

The input data does not specify true EV types and capacities. For each household, we derive an estimate of the respective "usable" EV capacity from the maximum amount charged in all observed transactions. These capacities range from 21.94 to 78.99kWh, with 7 out of 9 being between 35 to 49kWh. Subsequently, the SoC of the EV at the commencement of each transaction is the ratio of the total energy charged in the observed transaction to the derived usable EV capacity.
\begin{equation}
soc_t^{ev} = \frac{\sum_{t=i}^{t=j} d_{t}^{ev}}{soc_{max}^{ev}}
\end{equation}
where \( i \) is the starting point and \( j \) is the ending point of the charging transaction.

In this study, we assume an 11kW power rating and fully flexible load control for the EV charger, with fixed charging efficiency and loss. The assumptions further entail that users consistently use the same EV and provide the system with departure or disconnection times, for instance via an app. Similar to the EV capacity, we derive a “usable" PV peak from the maximum PV production recorded in 15-minute intervals. It is important to note that the true PV system and EV battery sizes are likely larger than these estimates which are based on net measurements and do not consider efficiency losses or partial capacity usage. Our profit calculation uses the average electricity purchase prices \citep{eurostat.2024} and feed-in rates \citep{Bundesnetzagentur.2024} from the year 2023. Lastly, we take the technical specifications of the BESS in this work  from the real solar batteries that are installed at the households in varying product generations. We specify technical details in Table \ref{tab:technicalspecifications}.

\begin{table}[h]

\centering
\caption{Technical \& Tariff Specifications}
\begin{tabular}{lll}
\toprule
\textbf{Definition}&  \textbf{Parameter}&\textbf{Value(s)}\\
\midrule
BESS inverter efficiency&  $\eta$&0.95\\
BESS standing loss&  $loss^{b}$&0.003\%\\
EV charger power&  $ev_{rate}^{max}$&11kW\\
Grid prices&  $p_{buy}$, $p_{sell}$&0.40€/kWh, 0.08€/kWh\\
\bottomrule
\end{tabular}
\label{tab:technicalspecifications}
\end{table}

\begin{figure}[H]
    \centering
    \includegraphics[width=0.6\linewidth]{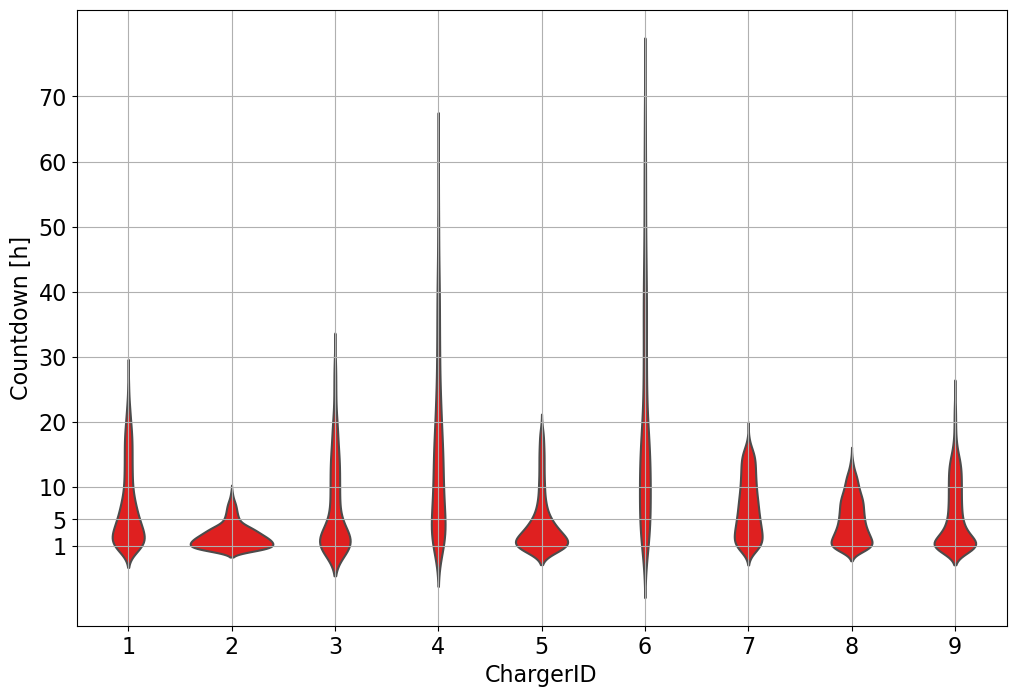}
    \caption{Distribution of countdown values}
    \label{fig:violin01to09}
\end{figure}

\begin{figure}[H]
    \centering

 \begin{subfigure}[b]{\textwidth}
        \centering
    \includegraphics[width=0.75\linewidth]{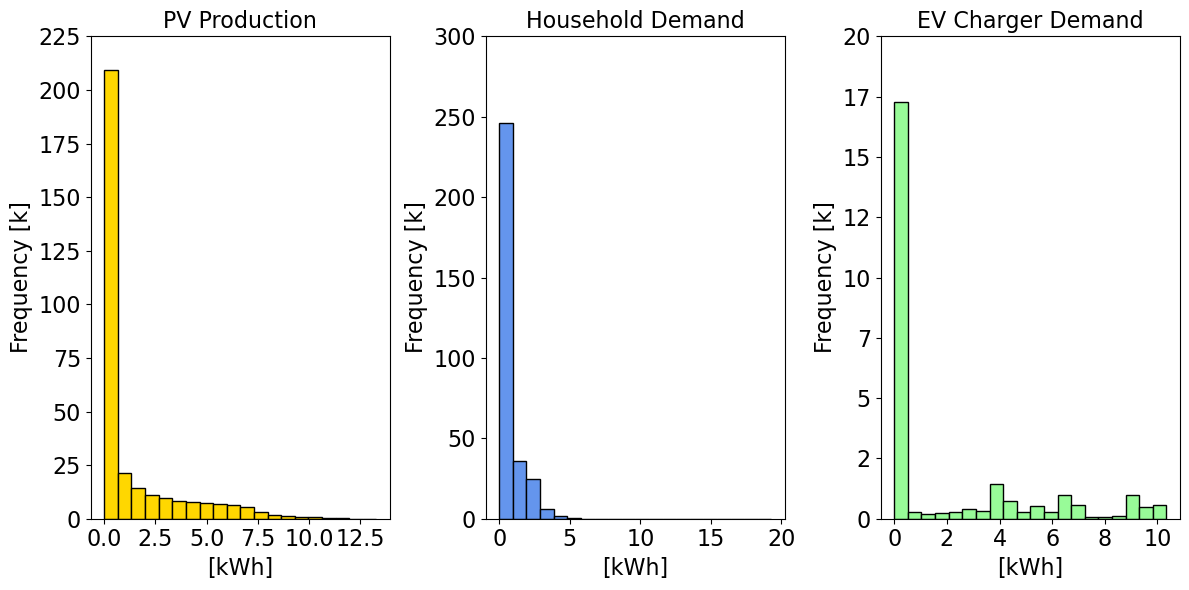}
    \end{subfigure}

 \begin{subfigure}[b]{\textwidth}
    \centering        \includegraphics[width=0.5\linewidth]{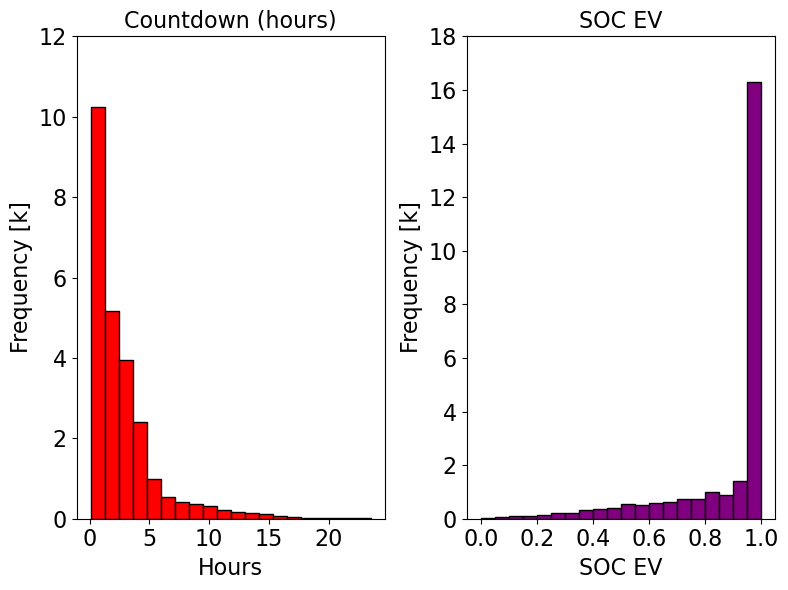}
    \end{subfigure}
    \caption{Histograms for HouseholdIDs}
    \label{fig:hist}
\end{figure}

In terms of EV connectivity and charging, HouseholdIDs 02 and 03 have the least EV connection time and the lowest number of charging transactions, while 09 stands out with the highest number of transactions. Interestingly, HouseholdID 06 has a similar total connection time to 09 but achieves this with fewer (longer) charging periods. The distribution of countdown values in the violin plots in Figure \ref{fig:violin01to09} indicate various duration frequencies. The plots reveal some transactions with long durations in HouseholdIDs 04 and 06, with most transactions being short (under 5 hours) or medium (5 to 15 hours). Users of HouseholdIDs 02 and 05 almost exclusively connect EVs for short periods. Additional histograms describe the distribution of the household features from Table \ref{tab:descriptiveAll} in Figure \ref{fig:hist}.

\begin{figure}[H]
    \centering
    \includegraphics[width=0.75\linewidth]{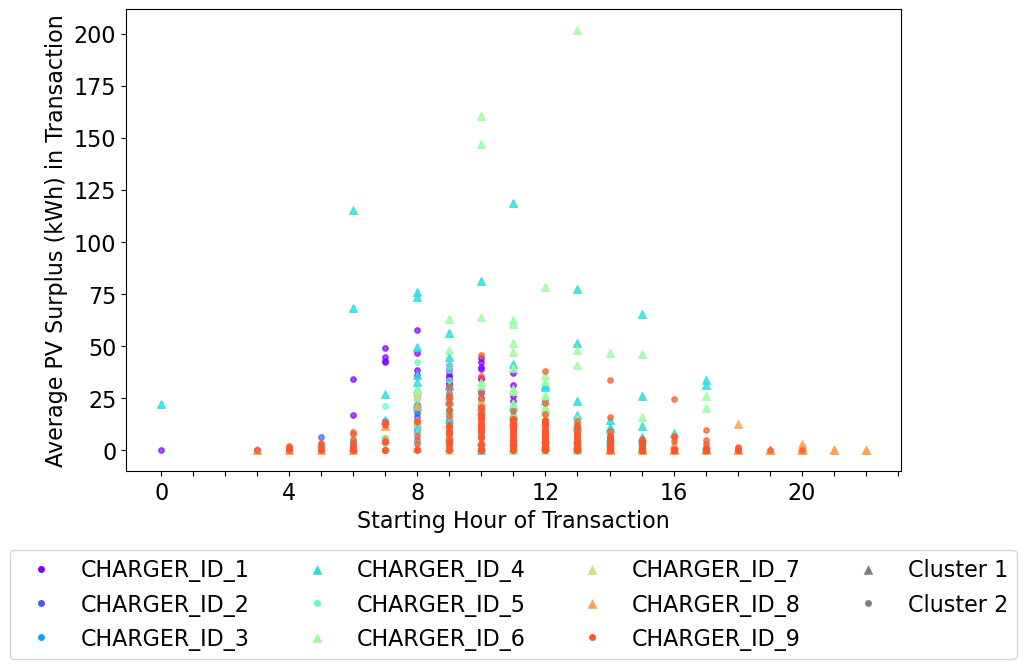}
    \caption{EV transaction start times and PV surplus}
    \label{fig:scatterTimes}
\end{figure}

For a deeper analysis of transaction start times in conjunction with PV surplus availability, we use a scatter plot in Figure \ref{fig:scatterTimes}. It includes each transaction observed in the nine datasets, its time of initiation, and the mean PV surplus recorded in the transaction period. This is useful in analyzing the timing of transactions that can be charged from PV power. It reveals that most transactions with PV surplus commence in typical PV production hours between 8:00 and 16:00. We note that charging EVs in power-mode would use the PV-surplus in these instances. Transactions that include PV surplus but start outside of these hours are infrequent and exclusively located in HouseholdIDs 01, 04, 06, and 09. Those are transactions in which power-mode charging would charge from the grid, despite PV-surplus being available later on.

\begin{figure}[H]
    \centering
    \includegraphics[width=0.75\linewidth]{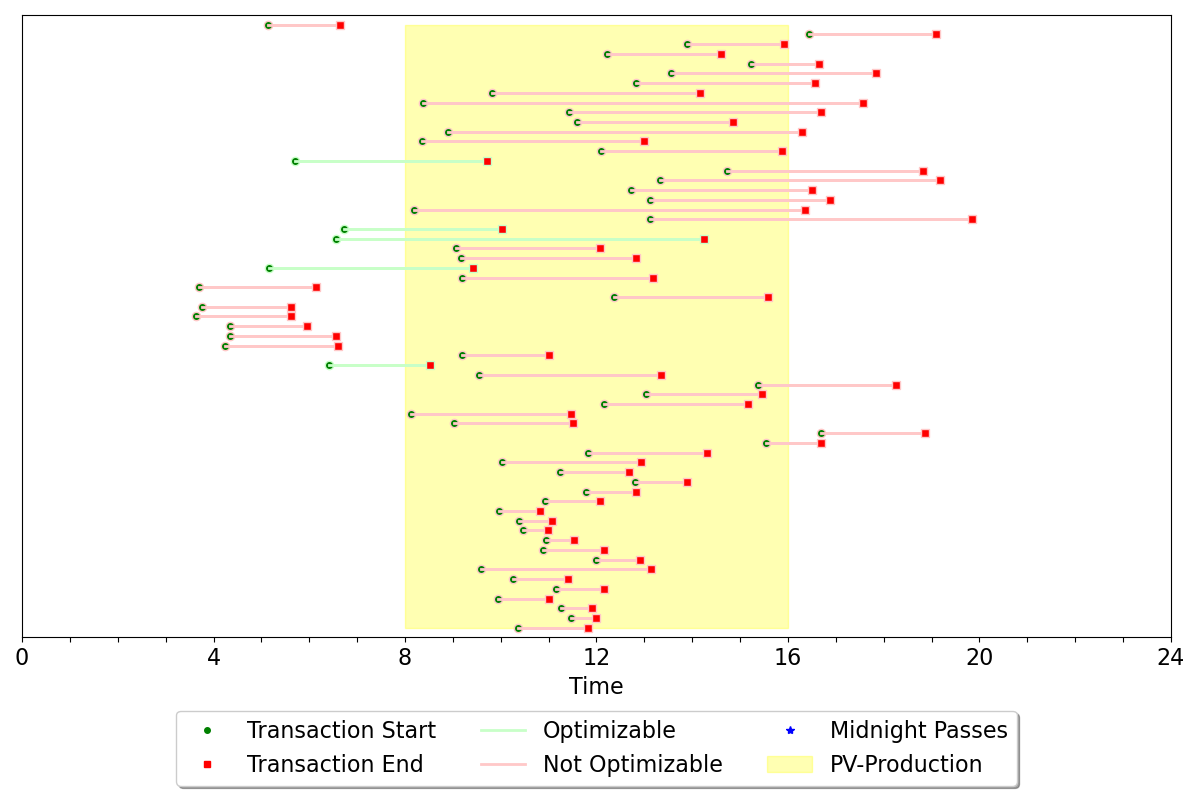}
    \caption{Timing of transactions and optimization potential (HouseholdID 02)}
    \label{fig:LineChart2}
\end{figure}

\begin{figure}[H]
    \centering
    \includegraphics[width=0.75\linewidth]{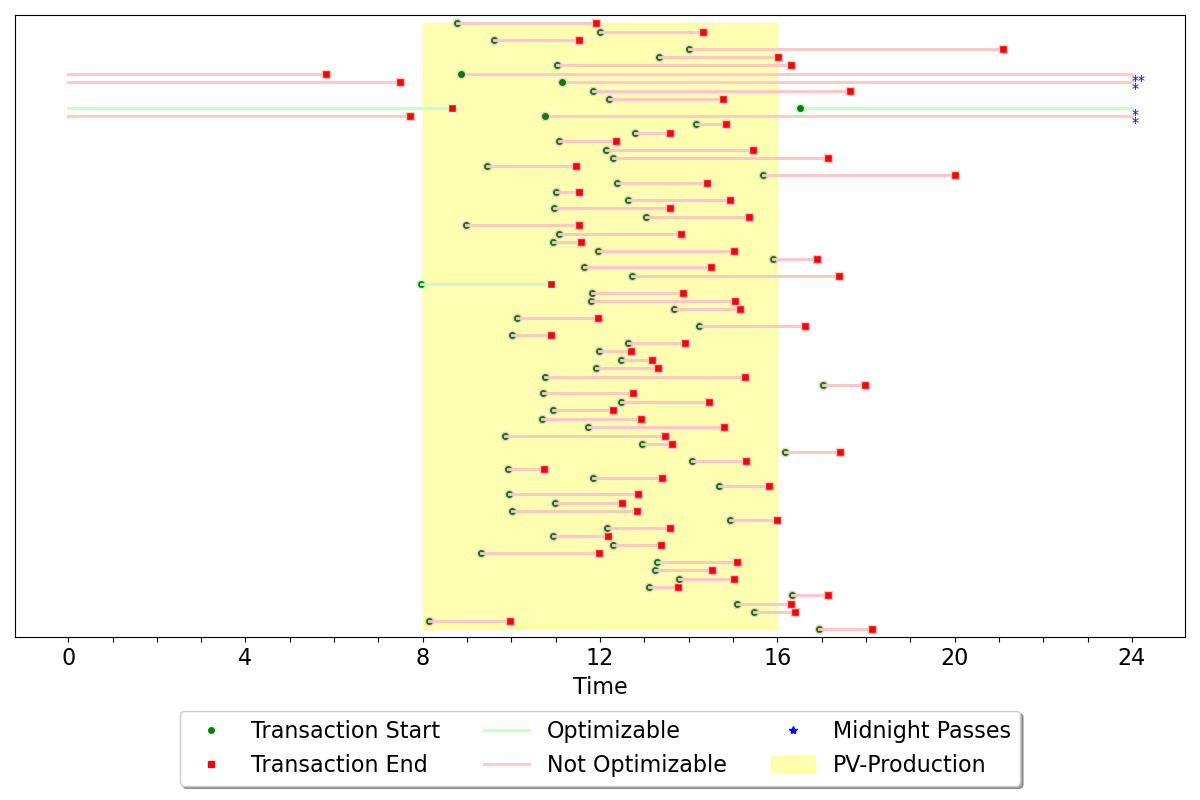}
    \caption{Timing of transactions and optimization potential (HouseholdID 03)}
    \label{fig:enter-label}
\end{figure}

\begin{figure}[H]
    \centering
    \includegraphics[width=0.75\linewidth]{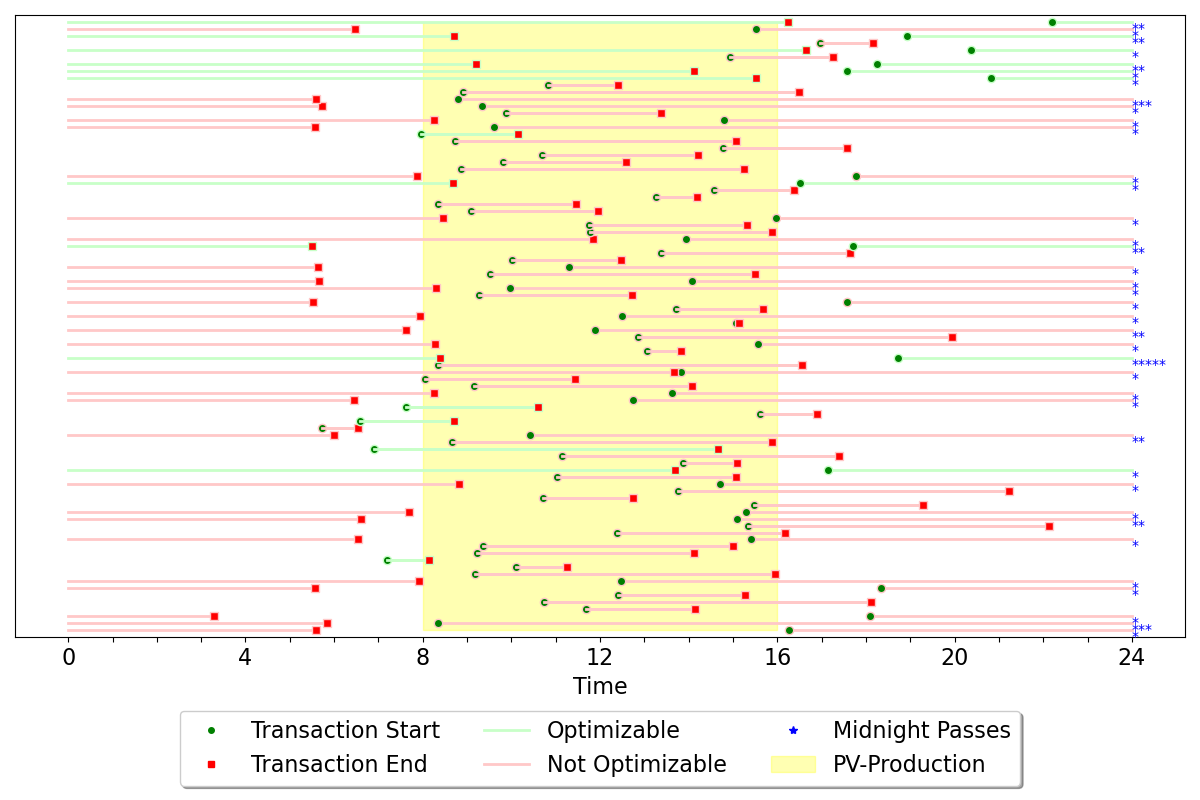}
    \caption{Timing of transactions and optimization potential (HouseholdID 04)}
    \label{fig:enter-label}
\end{figure}

\begin{figure}[H]
    \centering
    \includegraphics[width=0.75\linewidth]{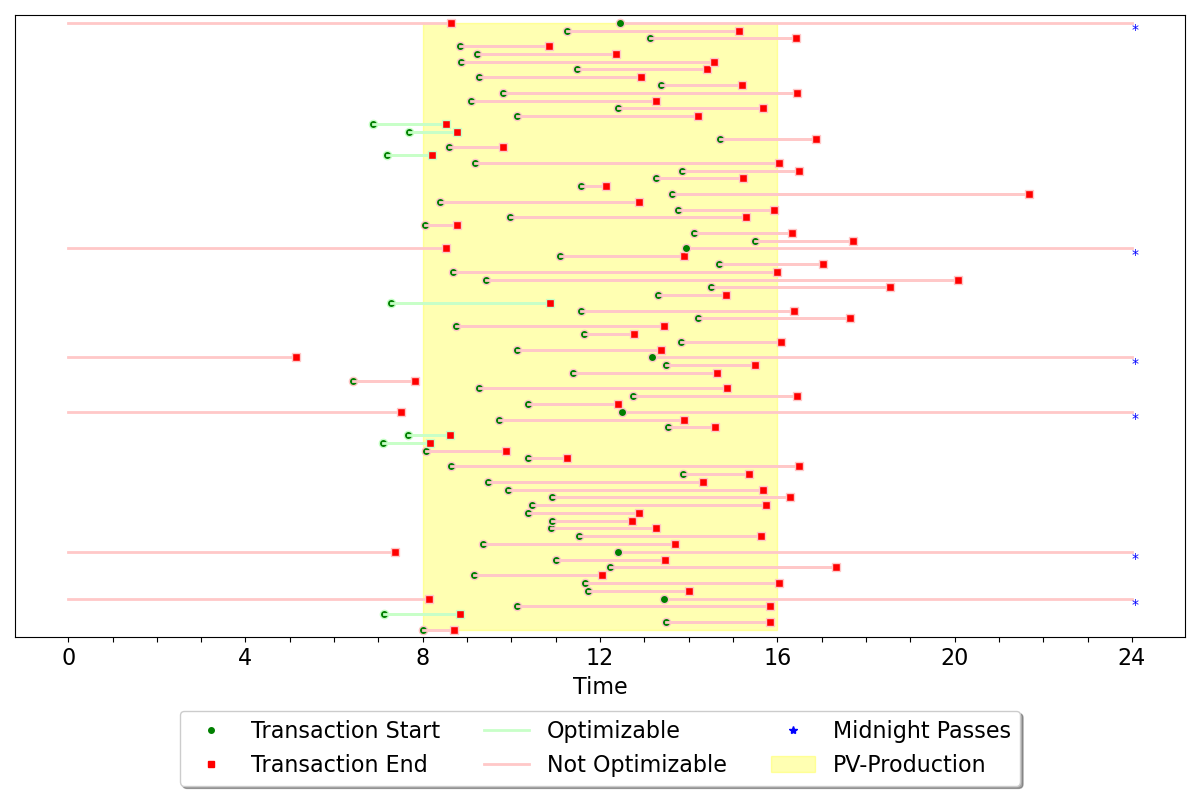}
    \caption{Timing of transactions and optimization potential (HouseholdID 05)}
    \label{fig:enter-label}
\end{figure}

\begin{figure}[H]
    \centering
    \includegraphics[width=0.75\linewidth]{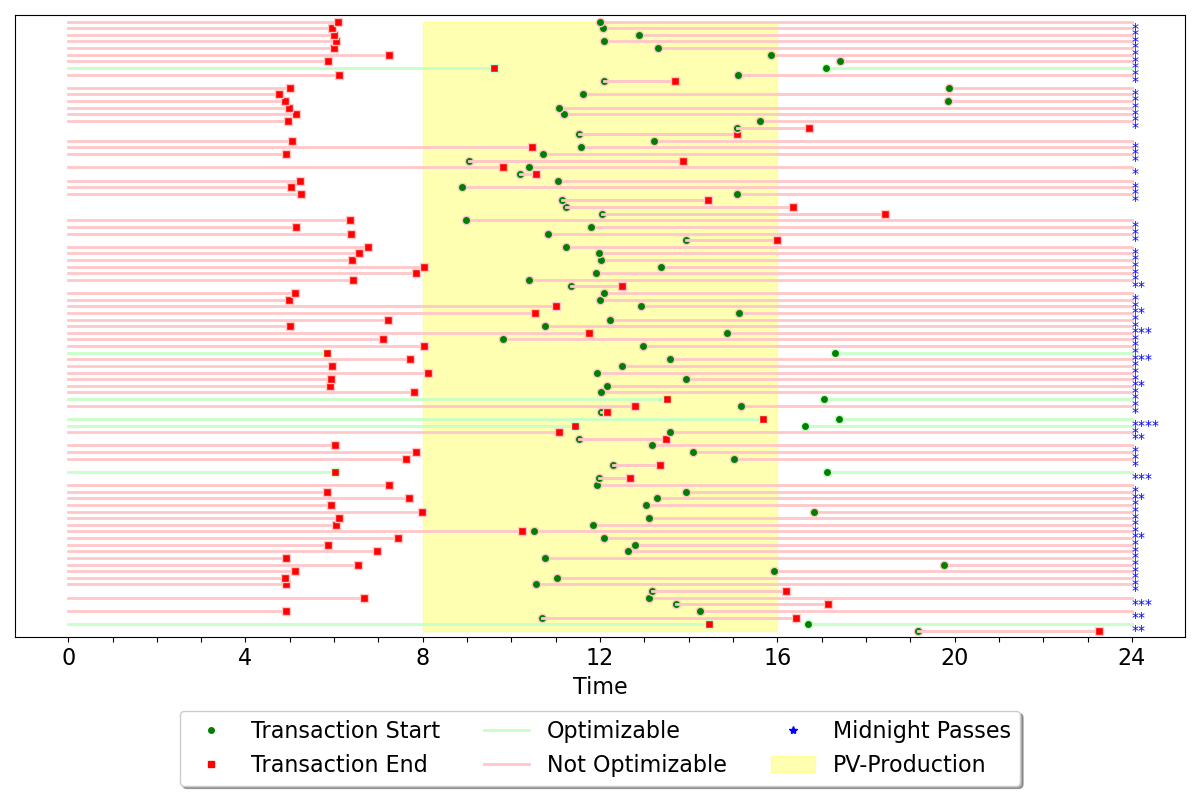}
    \caption{Timing of transactions and optimization potential (HouseholdID 06)}
    \label{fig:enter-label}
\end{figure}

\begin{figure}[H]
    \centering
    \includegraphics[width=0.75\linewidth]{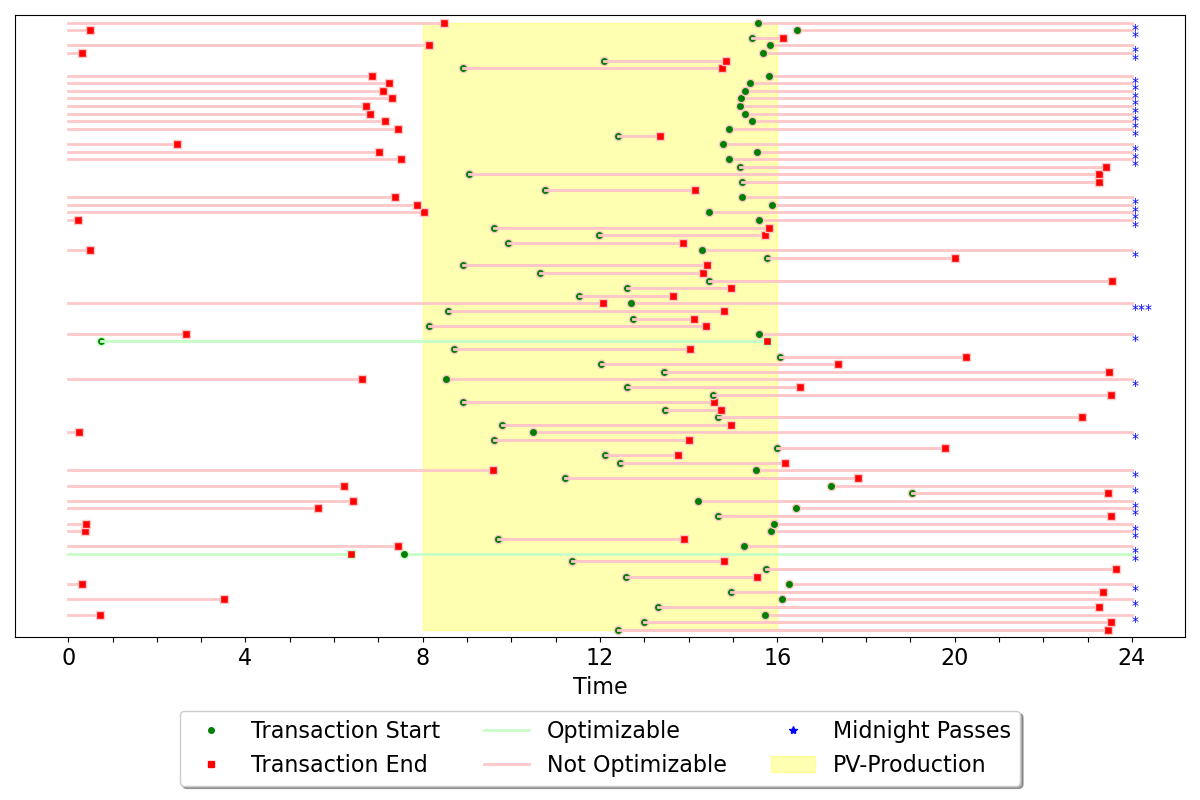}
    \caption{Timing of transactions and optimization potential (HouseholdID 07)}
    \label{fig:enter-label}
\end{figure}

\begin{figure}[H]
    \centering
    \includegraphics[width=0.75\linewidth]{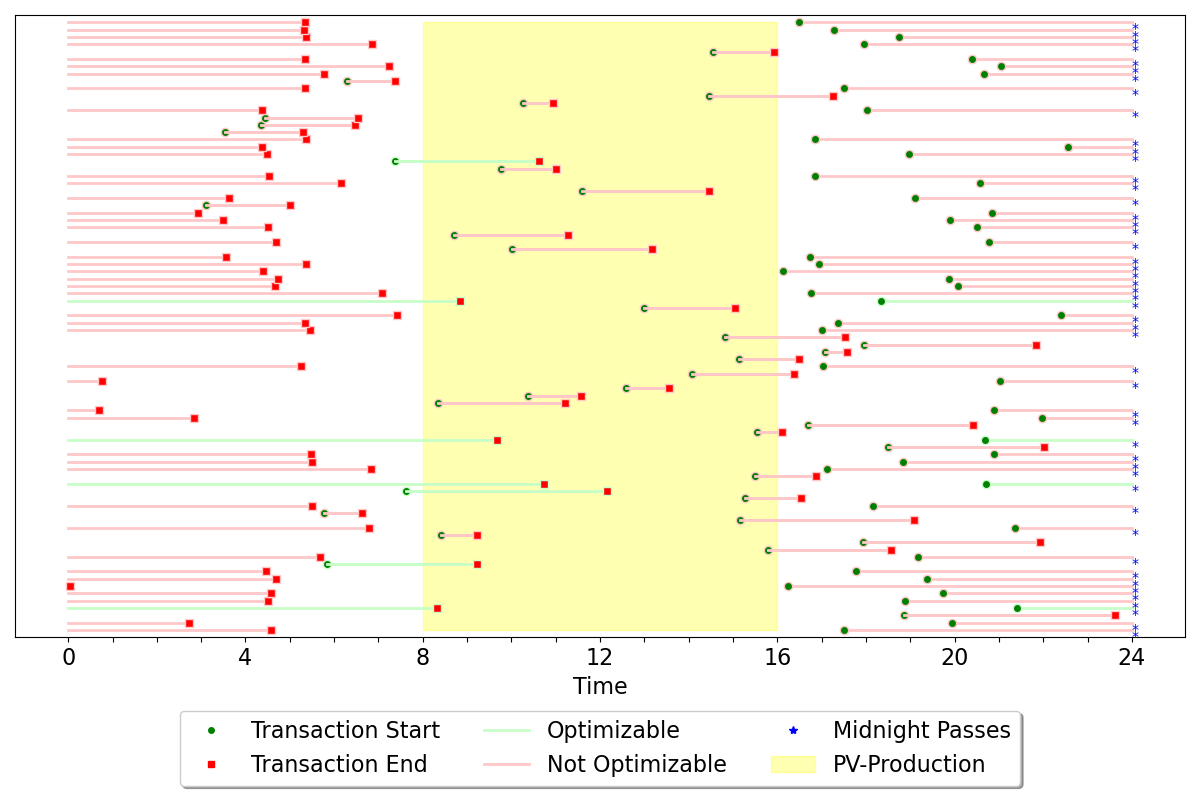}
    \caption{Timing of transactions and optimization potential (HouseholdID 08)}
    \label{fig:enter-label}
\end{figure}

\begin{figure}[H]
    \centering
    \includegraphics[width=0.75\linewidth]{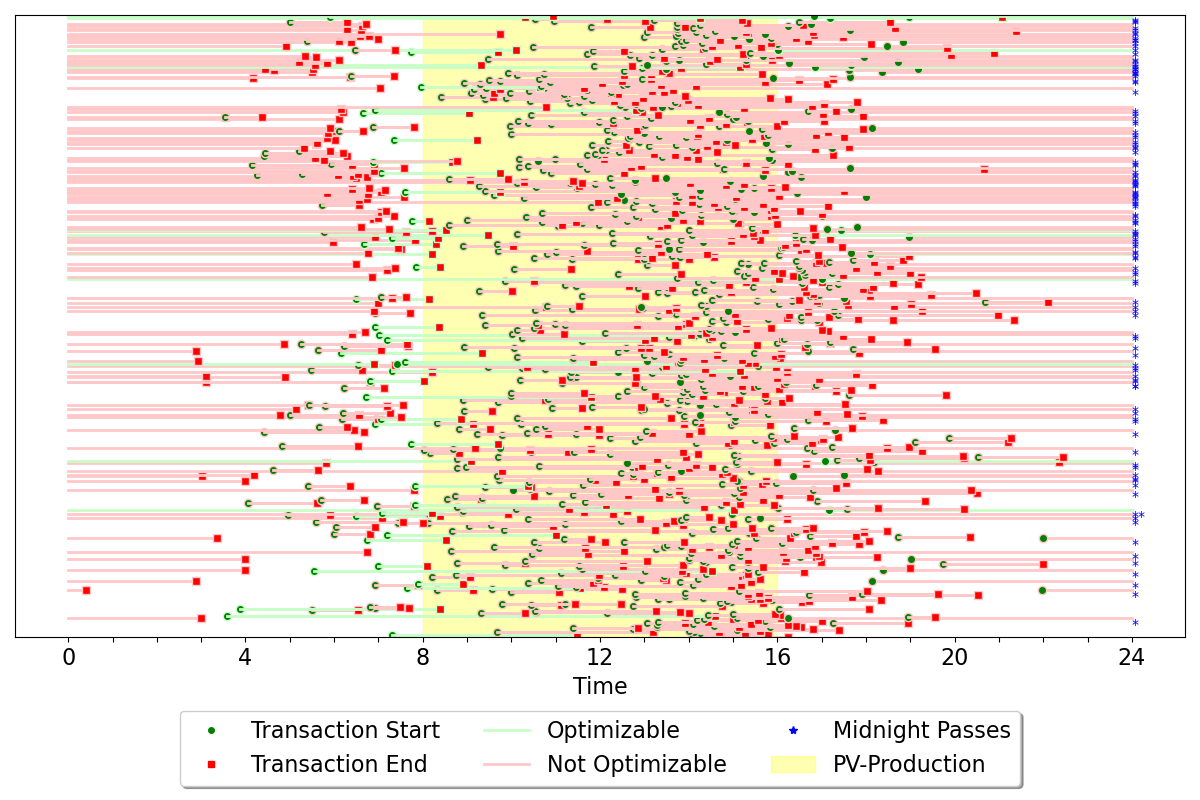}
    \caption{Timing of transactions and optimization potential (HouseholdID 09)}
    \label{fig:LineChart9}
\end{figure}

\begin{figure}[H]
    \centering
    \includegraphics[width=0.75\linewidth]{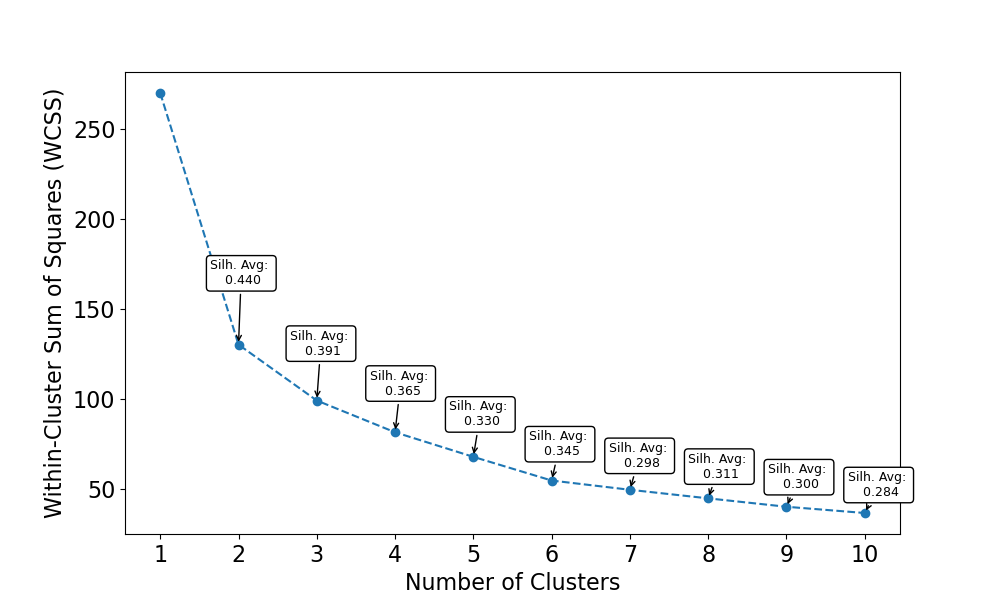}
    \caption{K-Means clustering (WCSS and Silhouette Scores)}
    \label{fig:elbow}
\end{figure}

\begin{figure}[H]
    \centering
    \includegraphics[width=0.9\linewidth]{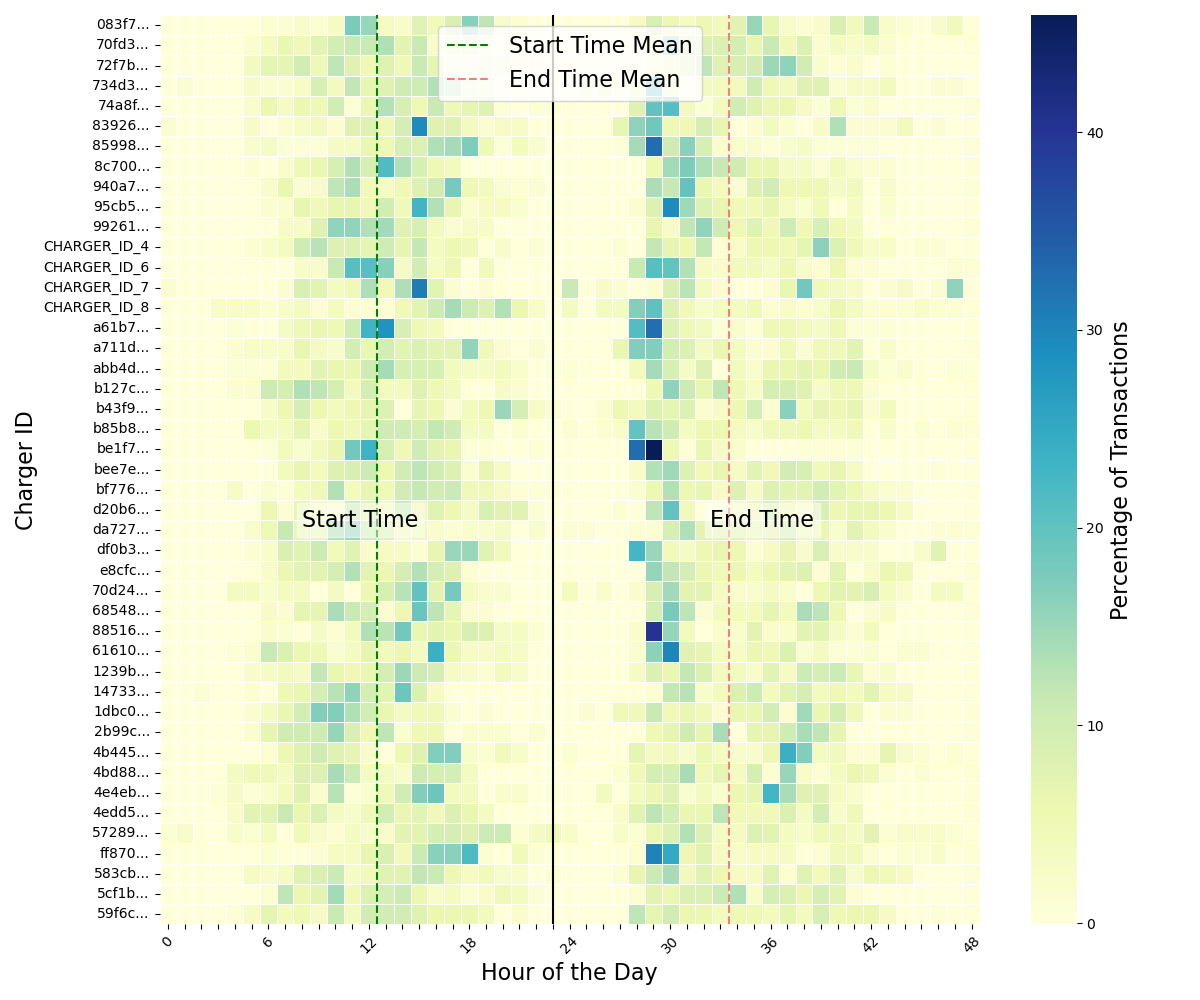}
    \caption{Heat map of start and end times of EV charging transactions for Cluster 1}
    \label{fig:cluster1}
\end{figure}

\begin{figure}[H]
    \centering
    \includegraphics[width=0.9\linewidth]{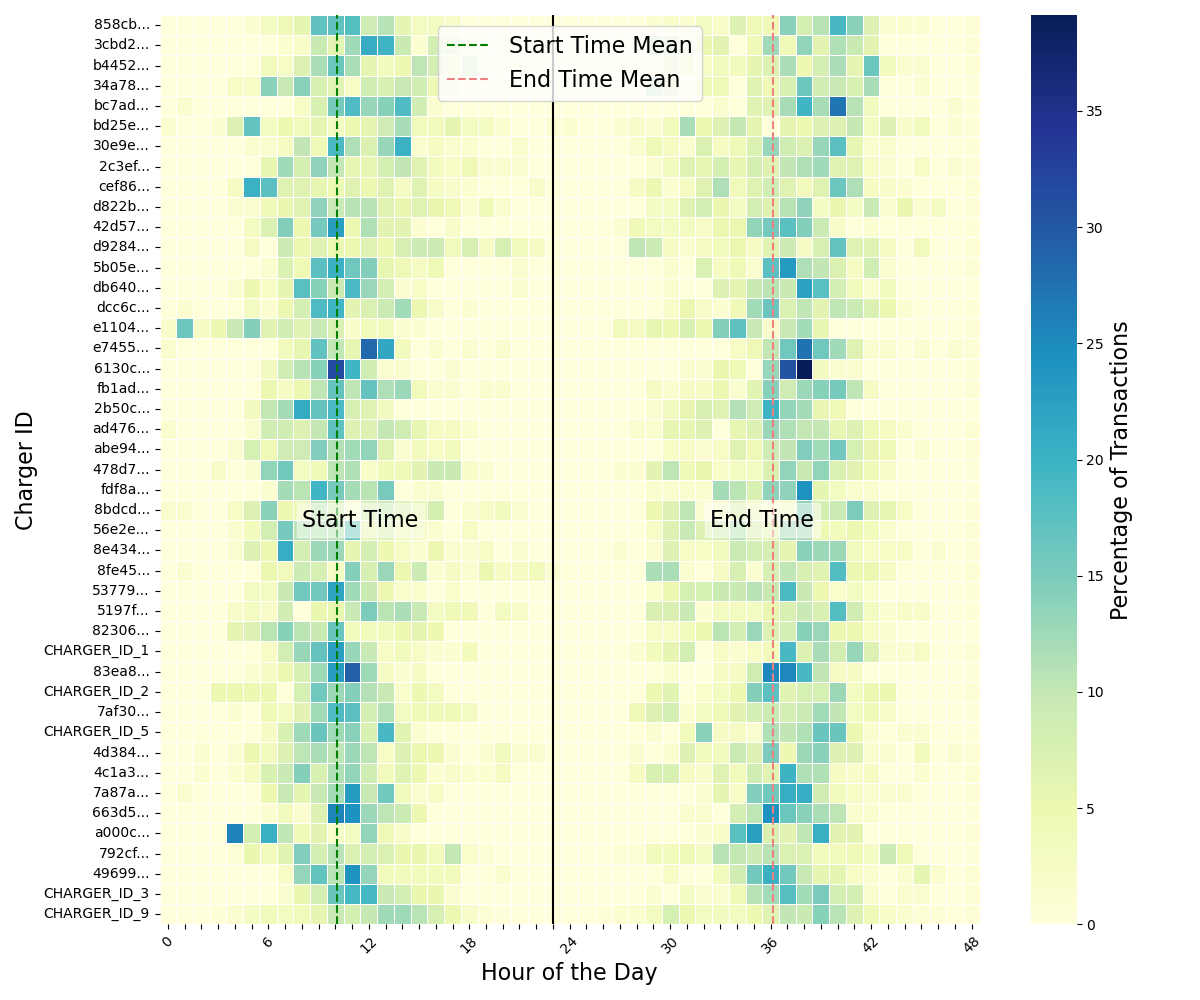}
    \caption{Heat map of start and end times of EV charging transactions for Cluster 2}
    \label{fig:cluster2}
\end{figure}

\begin{table}[H]
\centering
\caption{Potential Grid Savings}
\begin{tabular}{rrrrr}
\toprule
\textbf{Month}& \textbf{Housholds}& \textbf{Mean grid}& \textbf{StD grid}& \textbf{Mean grid}\\
 & & \textbf{savings}& \textbf{savings}&\textbf{savings}\\
 & \textbf{[n]}& \textbf{[Wh]}& \textbf{[Wh]}& \textbf{[\%]}\\
\midrule
11&   19&3,024.74& 7,833.01& 0.72\\
 12& 23& 535.96& 1,444.42&0.10\\
 01& 29& 1,044.48& 2,740.92&0.19\\
02&  36&3,208.31& 6,554.84& 0.73\\
03&  46&10,303.85& 12,762.11& 3.10\\
04&  51&11,697.71& 17,804.50& 5.58\\
05&  59&13,249.53& 18,424.27& 8.02\\
06&  69&8,544.61& 11,792.43& 8.20\\
07&  75&12,164.64& 17,793.93& 7.79\\
08&  82&9,758.50& 13,785.97& 6.65\\
 09& 82& 10,163.15& 13,762.46& 5.83\\
 10& 82& 10,424.74& 20,700.11& 3.16\\
\end{tabular}
\label{tab:gridsavings}
\end{table}

\section{Simulation Study}
\label{chap:simul}

To provide additional insights into our findings, we conduct a simulation study in three steps:
Step 1: Training, evaluation, and testing runs of our DDPG implementation on a synthetic dataset with higher optimization potential.
Step 2: Fine-tuning hyperparameters using an impact analysis and grid search.
Step 3: New training, evaluation, and testing of the DDPG model on the original data, using tuned hyperparameters.

\subsection{Synthetic Dataset}

In our simulation study, HouseholdID 06, which offers a balance of EV charging frequency and durations, serves as a base to derive a synthetic dataset. We double the EV charging transactions and shift the connection time to start before typical PV surplus times. To address the shortages of PV surplus, we increase PV production by a factor of 1.5. To limit the impact of the BESS as a buffer for inefficient EV charging this study includes the smallest BESS model (6.75kWh capacity and 3.3kW inverter power). We train, evaluate, and test in the same way as for the datasets HouseholdID 01 to 09.

\begin{figure}[H]
    \centering
    \includegraphics[width=1\linewidth]{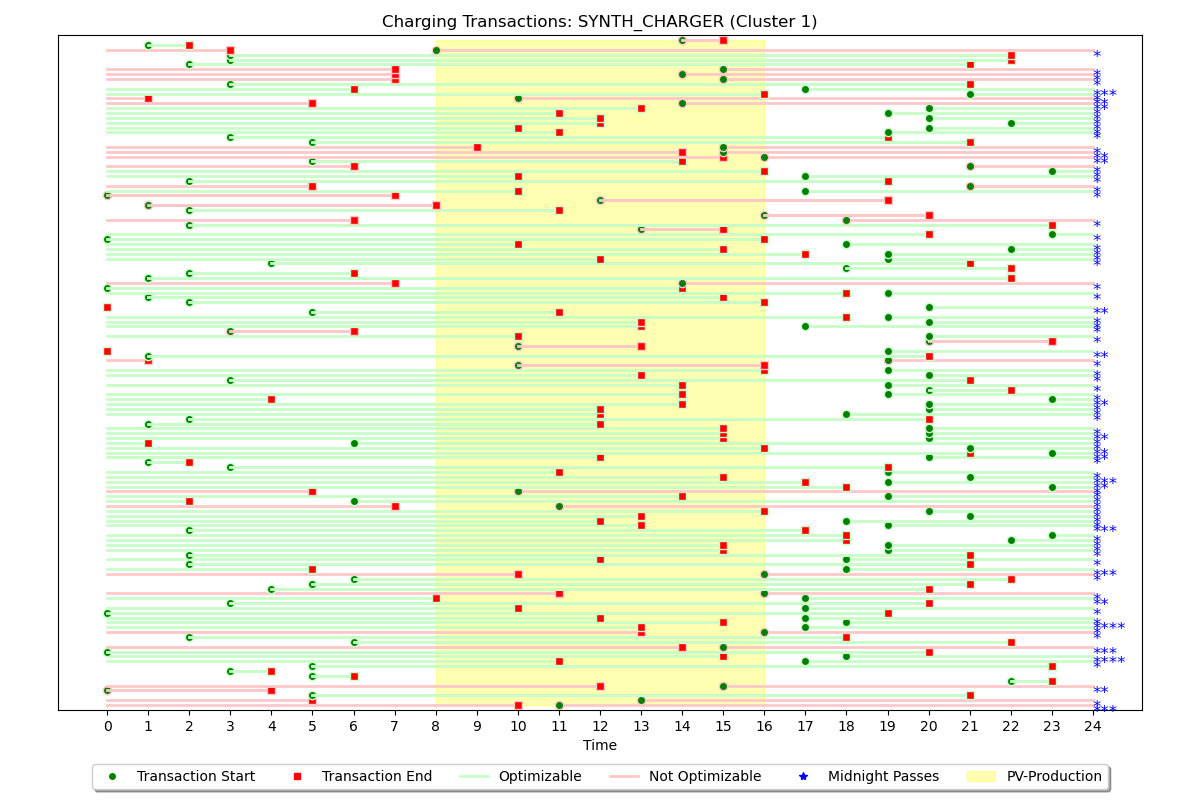}
    \caption{Timing of transactions and optimization potential (Synthetic Dataset)}
    \label{fig:LineChartSynth}
\end{figure}

Figure \ref{fig:LineChartSynth} applies the synthetic data to the line chart from  \ref{chap:descriptiveanalysis} to illustrate the impact of these changes on the frequency and timing of EV charging transactions. The lines indicate a high frequency of EV connection times with optimization potential, transactions in which power-mode EV charging is inefficient.

\subsubsection{Hyperparameter Tuning}

Utilizing our synthetic dataset, we tune hyperparameters in multiple steps: An impact analysis and a two-step grid search. Because the model is computationally expensive, the run time of training and evaluating multiple versions for tuning presents a challenge. This necessitates limiting the tuning to a careful selection of parameters. Since the original parameters from \cite{Langer.2022} prove effective in previous studies by \cite{Fujimoto.2018, Lillicrap.2019, Raffin.2021, Yu.2021} we focus on parameters related to the specialization of this work. Particularly, we identify parameters affecting the rate of exploration and the parameters related to our particular construction of the reward function.

The first step involves assigning two alternative values to each parameter in Table \ref{tab:parametersearch} individually. Each of the resulting 16 models runs 40 times on varying random seeds. We select the four parameters with the highest impact on results for the grid search using the values in Table \ref{tab:gridsearch}. Grid search explores any possible combinations of these values resulting in a total of 81 models  ($3^4$). To manage computational intensity and long run times the grid search trains and evaluates 10 random seeds in the first step, before running the full 40 random seeds on the four most promising models. Finally, the tuned DDPG model trains, evaluates, and tests anew on the nine original datasets.

Interestingly, our results on the synthetic data set show that the original settings prove reliable in producing the best results, except for a small improvement when reducing the DNN sizes of the actor and critic networks from 300 to 250 nodes in layer one and from 600 to 500 nodes in layer two.

\begin{table}
\centering
\caption{Parameter Search}
\resizebox{\textwidth}{!}{%
\begin{tabular}{ll|ll}
\toprule
\textbf{Name and symbol} & \textbf{Original Value}& \textbf{Alternative 1}&\textbf{Alternative 2}\\
Batch size K & 120  & 100&150\\
 Buffer size N & 24,000 & 20,000&30,000\\
Learning rates actor  $\alpha_a$ \& $\alpha_c$& 0.0001; 0.001& 0.0005; 0.005&0.00005 ; 0.0005\\
Exploration noise $\mathscr{N}$& $\mathscr{N}(0, 0.1)$ & $\mathscr{N}(0, 0.2)$&Ornstein-Uhlenbeck-Noise\\
 Soft update rate $\tau$ &  0.001  & 0.005&0.0005\\
 Discomfort weight& 0.01 & 0.04&linear discomfort function\\
 Neural network size L & 300; 600& 200; 400&400; 800\\
 Penalty weight& 0.01& 0.00&1.00\\
\end{tabular}
}
\label{tab:parametersearch}
\end{table}

\begin{table}
\centering
\caption{Grid Search }
\resizebox{\textwidth}{!}{%
\begin{tabular}{ll|ll}
\toprule
\textbf{Name and symbol} & \textbf{Original Value}& \textbf{Alternative 1}&\textbf{Alternative 2}\\
Batch size K & 120  & 100&150\\
Learning rates actor  $\alpha_a$ \& $\alpha_c$& 0.0001; 0.001& 0.0005; 0.005&0.00005 ; 0.0005\\
Exploration noise $\mathscr{N}$& $\mathscr{N}(0, 0.1)$ & $\mathscr{N}(0, 0.2)$&$\mathscr{N}(0, 0.4)$\\
 Neural network size L & 300; 600& 200; 400&\textbf{250; 500}\\
\end{tabular}
}
\label{tab:gridsearch}
\end{table}

\subsection{Tuned Model Performance}

To conclude the simulation study, we report the testing results of the tuned DDPG-DRL model on the synthetic dataset in the first line of Table \ref{tab:ResultsFinTest} and the tuning results on HouseholdID 01 to 09 in the following lines. We report the respective evaluation results in Table \ref{tab:ResultsFinEval}. For compatibility, the result tables mirror those in Chapter \ref{chap:modelresults}, noting that benchmark results for the original datasets remain unchanged.

\begin{table}
\centering
\caption{Tuned Model - Testing Results (125 Days)}
\resizebox{\textwidth}{!}{%
\begin{tabular}{c|ccc|ll|ll|ll}
\toprule
& \multicolumn{2}{c}{\textbf{Benchmarks}}&& \multicolumn{5}{c}{\textbf{DRL}} &\\
 & Lower& Upper& Optimization& \multicolumn{2}{c}{Result}& \multicolumn{2}{c}{Potential Realized}&\multicolumn{2}{c}{Disco}\\
 \textbf{HouseholdID}& (RBPM)& (MPC)& Potential& Mean& Best Eval& Mean& Best Eval&Mean&Best Eval\\

\midrule

 &\multicolumn{5}{c}{Mean profit from grid interaction [€/day]}&&&Pp.&\\
\midrule
 SYNTHETIC& -4.09& -2.96& 1.13& -3.28& -3.22& 71\%& 76\%&4.81&2.88\\
\midrule
HouseholdID 01& 0.22& 0.49& 0.27& 0.31& 0.38& 34\%& 61\%&1.42 &0.45\\
HouseholdID 02& -2.24& -2.23& 0.01& -2.30& -2.24& 0\%& 0\%&1.48 &1.43\\
HouseholdID 03& -0.83& -0.76& 0.07& -1.01& -0.92& 0\%& 0\%&2.86 &1.86\\
HouseholdID 04& -0.32& -0.12& 0.20& -0.42& -0.32& 0\%& 0\%&4.28 &3.12\\
HouseholdID 05& -0.16& -0.05& 0.11& -0.20& -0.15& 0\%& 12\%&4.62 &3.67\\
HouseholdID 06& -5.96& -5.87& 0.09& -6.09& -5.99& 0\%& 0\%&4.97 &7.07\\
 HouseholdID 07& -5.62& -5.61& 0.01& -5.66& -5.63& 0\%& 0\%&1.64 &3.45\\
 HouseholdID 08& -3.60& -3.56&0.04& -3.66& -3.61& 0\%& 0\%&9.73 &2.22\\
 HouseholdID 09& -4.17& -3.82& 0.36& -4.03& -3.94& 41\%& 65\%&5.21 &1.00\\
\end{tabular}
}
\label{tab:ResultsFinTest}
\end{table}

\begin{table}[H]
\centering
\caption{Tuned Model - Evaluation Results (60 Days)}
\resizebox{\textwidth}{!}{%
\begin{tabular}{c|ccc|ll|ll|ll}
\toprule
& \multicolumn{2}{c}{\textbf{Benchmarks}}&& \multicolumn{5}{c}{\textbf{DRL}} &\\
 & Lower& Upper& Optimization& \multicolumn{2}{c}{Result}& \multicolumn{2}{c}{Potential Realized}&\multicolumn{2}{c}{Disco}\\
 \textbf{HouseholdID}& (RBPM)& (MPC)& Potential& Mean& Best Eval& Mean& Best Eval&Mean&Best Eval\\
 
\midrule

 &\multicolumn{5}{c}{Mean profit from grid interaction [€/day]}&&&Pp.&\\
\midrule
 SYNTHETIC& -4.87& -3.20& 1.68& -3.46& -3.29& 84\%& 95\%&2.39&2.63\\
\midrule
HouseholdID 01& & 0.48& 0.36& 0.29& 0.42& 46\%& 82\%&1.76&0.95\\
HouseholdID 02& -2.37& -2.36& 0.01& -2.44& -2.37& 0\%& 15\%&1.84&1.39\\
HouseholdID 03& -1.25& -1.24& 0.01& -1.33& -1.26& 0\%& 0\%&8.87&0.32\\
HouseholdID 04& -1.20& -1.02& 0.18& -1.29& -1.16& 0\%& 22\%&3.10&2.34\\
HouseholdID 05& -0.27& -0.17& 0.10& -0.39& -0.25& 0\%& 19\%&2.02&3.85\\
HouseholdID 06& -6.09& -5.81& 0.29& -6.04& -5.89& 19\%& 70\%&5.02&1.64\\
 HouseholdID 07& -5.16& -5.16& 0.00& -5.21& -5.16& 0\%& 0\%&5.96&7.86\\
 HouseholdID 08& -3.92& -3.91&0.02& -3.99& -3.92& 0\%& 23\%&3.94&2.41\\
 HouseholdID 09& -4.64& -4.22& 0.42& -4.43& -4.32& 51\%& 77\%&1.12&0.17\\
\end{tabular}
}
\label{tab:ResultsFinEval}
\end{table}

As expected, the changes to the input data lead to a much higher optimization potential at 1.13€/day in the synthetic dataset, more than triple the optimization potential of any of the original datasets. Given this input, the tuned DDPG model can on average realize over 70\% of the optimization potential, 76\% when using the agent with the best evaluation results. The results of the DRL model confirm our findings that the DRL-DDPG performance relies on sufficient optimization potential in the input data.

Using the tuned version of the model with reduced actor and critic network sizes, we record slight improvements of 0.01 to 0.07€/day in the results for all datasets except HouseholdID 08. In this special case, the results were 0.02-0.03€/day lower. The best-eval agents for HouseholdID 01 and 09 now realize 61\% and 65\% of the optimization potential. This constitutes an improvement of 15 and 7 percentage points. HouseholdID 05's best-eval agent now beats the lower benchmark marginally. This does, however, not change the overall picture of the DRL performance. In conclusion, our results show that the DRL-DDPG model excels on high optimization potential data, and nearly matches the near-optimal RBPM benchmark in the absence of optimization potential. It falters when faced with low optimization potential.







 \bibliographystyle{elsarticle-num} 
 \bibliography{references}







\end{document}